\documentclass[fleqn,usenatbib]{mnras}

\usepackage{newtxtext,newtxmath}

\usepackage[T1]{fontenc}

\DeclareRobustCommand{\VAN}[3]{#2}
\let\VANthebibliography\thebibliography
\def\thebibliography{\DeclareRobustCommand{\VAN}[3]{##3}\VANthebibliography}

\usepackage{graphicx}	
\graphicspath{{Images/}}
\usepackage{amsmath}	
\usepackage{ulem}
\usepackage[dvipsnames]{xcolor}
\usepackage{multirow}

\newcommand{\Halpha}{\ifmmode {\rm H}\alpha \else H$\alpha$\fi}
\newcommand{\Hbeta}{\ifmmode {\rm H}\beta \else H$\beta$\fi}
\newcommand{\Hgamma}{\ifmmode {\rm H}\gamma \else H$\gamma$\fi}
\newcommand{\Hdelta}{\ifmmode {\rm H}\delta \else H$\delta$\fi}
\newcommand{\Lya}{\ifmmode {\rm Ly}\alpha \else Ly$\alpha$\fi}
\newcommand{\Lyb}{\ifmmode {\rm Ly}\beta \else Ly$\beta$\fi}
\newcommand{\HeI}{\ifmmode {\rm He}\,\textsc{i}\,\lambda5876 \else 
                  He\,\textsc{i}\,$\lambda5876$\fi}
\newcommand{\HeII}{\ifmmode {\rm He}\,\textsc{ii}\,\lambda4686 \else 
                   He\,\textsc{ii}\,$\lambda4686$\fi}

\newcommand{\mgii}{Mg\,\textsc{ii}}
\newcommand{\nev}{[Ne\,\textsc{v}]}
\newcommand{\Nev}{[Ne\,\textsc{v}]$3426\AA$}
\newcommand{\civ}{\ifmmode {\rm C}\,\textsc{iv} \else C\,\textsc{iv}\fi}
\newcommand{\Civ}{C\,\textsc{iv}$1549\AA$}

\newcommand{\kms}{\hbox{km$\,$s$^{-1}$}}
\newcommand{\nh}{N_{\text{H}}}

\newcommand{\eddrat}{\ifmmode{\lambda_{\text{Edd}}} \else  $\lambda_{\text{Edd}}$ \fi}
\newcommand{\mbh}{\ifmmode{M_{\text{BH}}} \else  $M_{\text{BH}}$ \fi}
\newcommand{\simba}{\textsc{Simba}}
\title[Tracing AGN-Galaxy Co-Evolution with UV Line-Selected Obscured AGN]{Tracing AGN-Galaxy Co-Evolution with UV Line-Selected Obscured AGN}

\author[L. Barchiesi et al.]{
Luigi Barchiesi,$^{1,2,3}$\thanks{E-mail: luigi.barchiesi@uct.ac.za}
L. Marchetti,$^{1,2,3}$
M. Vaccari,$^{2,3,4}$
C. Vignali,$^{5,6}$
F. Pozzi,$^{5,6}$
I. Prandoni,$^{3}$
R. Gilli,$^{6}$
\newauthor M. Mignoli,$^{6}$
J. Afonso,$^{7}$
V. Singh,$^{8}$
C. L. Hale,$^{9,10}$
I. Heywood,$^{9,11,12}$
M.J. Jarvis,$^{9,13}$
I. H. Whittam.$^{9,13}$
\\
$^{1}$Department of Astronomy, University of Cape Town, 7701 Rondebosch, Cape Town, South Africa\\
$^{2}$Inter-University Institute for Data Intensive Astronomy, Department of Astronomy, University of Cape Town, 7701 Rondebosch,
Cape Town, South Africa\\
$^{3}$INAF - Istituto di Radioastronomia, via Gobetti 101, 40129 Bologna, Italy\\
$^{4}$Inter-University Institute for Data Intensive Astronomy, Department of Physics and Astronomy, University of the Western Cape,
7535 Bellville, Cape Town, South Africa\\
$^{5}$Dipartimento di Fisica e Astronomia ``Augusto Righi'', Universit\`a degli Studi di Bologna, via P. Gobetti 93/2, 40129 Bologna, Italy\\
$^{6}$INAF-OAS, Osservatorio di Astrofisica e Scienza dello Spazio di Bologna, via Gobetti 93/3, 40129 Bologna, Italy\\
$^{7}$Instituto de Astrof\'{i}s\'{i}ca e Ci\^{e}ncias do Espa\c{c}o, Faculdade de Ci\^{e}ncias, Universidade de Lisboa, OAL, Tapada da Ajuda, PT1349-018
Lisboa, Portugal\\
$^{8}$ Astronomy and Astrophysics division, Physical Research Laboratory, Ahmedabad, 380009, Gujarat, India\\
$^{9}$ Astrophysics, Department of Physics, University of Oxford, Keble Road, Oxford, OX1 3RH, UK\\
$^{10}$ Institute for Astronomy, University of Edinburgh, Royal Observatory, Blackford Hill, Edinburgh, EH9 3HJ, UK \\
$^{11}$Department of Physics and Electronics, Rhodes University, PO Box 94, Makhanda, 6140, South Africa\\
$^{12}$South African Radio Astronomy Observatory, 2 Fir Street, Black River Park, Observatory, Cape Town 7925, South Africa\\
$^{13}$ Department of Physics and Astronomy, University of the Western Cape, Robert Sobukwe Road, 7535 Bellville, Cape Town, South Africa\\
}
\date{Accepted 2025 November 14. Received 2025 November 04; in original form 2025 September 02}

\pubyear{\the\year{}}

\begin{document}
\label{firstpage}
\pagerange{\pageref{firstpage}--\pageref{lastpage}}
\maketitle

\begin{abstract}
Understanding black hole–galaxy co-evolution and the role of AGN feedback requires complete AGN samples, including heavily obscured systems. Such sources are key to constraining the black hole accretion rate density over cosmic time, yet they are challenging to identify and characterise across most wavelengths. In this work, we present the first UV line–selected (\Nev\ and \Civ) sample of obscured AGN with full X-ray-to-radio coverage, assembled by combining data from the Chandra COSMOS Legacy survey, the COSMOS2020 UV–NIR catalogue, mid- and far-IR photometry from XID+, and radio observations from the VLA and MIGHTEE surveys. Using CIGALE to perform spectral energy distribution (SED) fitting, we analyse 184 obscured AGN at $0.6 < z < 1.2$ and $1.5 < z < 3.1$, enabling detailed measurements of AGN and host galaxy properties, and direct comparison with \simba\ hydrodynamical simulations. We find that X-ray and radio data are essential for accurate SED fits, with the radio band proving critical when X-ray detections are missing or in cases of poor IR coverage. Comparisons with matched non-active galaxies and simulations suggest that the \nev-selected sources are in a pre-quenching stage, while the \civ-selected ones are likely quenched by AGN activity. Our results indicate that \nev\ and \civ\ selections target galaxies in a transient phase of their co-evolution, characterised by intense, obscured accretion, and pave the way for future extensions with upcoming large area high-$z$ spectroscopic surveys.
\end{abstract}

\begin{keywords}
galaxies: active -- galaxies: evolution -- radio continuum: galaxies -- galaxies: emission lines
\end{keywords}



\section{Introduction}\label{sec:introduction}

It is widely accepted that the evolution of galaxies is closely intertwined with that of supermassive black holes (SMBH). Likewise, the growth of SMBHs appears to be influenced by the physical properties and evolutionary stages of their host galaxies. However, the exact driving mechanism of this co-evolution, as well as its timescales, is still not well understood. The discovery of scaling relations between SMBH mass and host-galaxy properties—such as stellar mass, stellar velocity dispersion, and bulge mass \citep[e.g.,][]{kormendy95,magorrian98,ferrarese00,gebhardt00,kormendy13}—provided the first strong evidence for a connection between black hole growth and galaxy evolution. These correlations led to the formulation of the co-evolution paradigm, which suggests that the assembly of galaxies and their central SMBHs are linked \citep[e.g.,][]{hopkins06,lapi14}. In its simplest form, this paradigm states that factors, both external \citep[such as a galaxy mergers, ][]{silk98,dimatteo05,treister12,lamastra13} and internal \citep[e.g.,][]{lapi14}, set off an intense phase of star-formation (SF), which, in turns, funnel a fraction of the gas into the central region of the galaxy, where it accretes into the SMBH and triggers the Active Galactic Nuclei (AGN) activity. In this first phase, the core of the galaxy is likely obscured due to the large quantity of gas and dust around the SMBH \citep{satyapal14,ellison19,secrest20}. The feedback from the now active galactic nucleus removes, heats up, and expels the gas on galaxy scales, effectively quenching the SF \citep[e.g.][]{zana22}. At the same time, gas is becoming scarce in the core region, and the AGN transitions to a type 1, i.e. unobscured, AGN -  if the line of sight is adequate. When there is no more gas available, the AGN activity ends, and the galaxy continues on as a ``red and dead'' elliptical galaxy \citep[e.g.,][]{hopkins07,cattaneo09}.\par
While there are a multitude of different variations on the co-evolution paradigm, almost all scenarios agree that a key phase is represented by the obscured accretion phase, where the high quantity of gas, fuelling the SF and the AGN accretion, also hides the AGN activity. This phase is of crucial importance in the co-evolution scenarios; however, it is also the least studied, mostly due to the challenge in detecting and recognising such obscured AGN \citep[e.g.,][]{rowan-robinson97,hughes98,martinez-sansigre05}. \par
Additionally, very obscured AGN are required to reproduce the X-ray Background (XRB) spectrum. The integrated contribution of all known AGN, obtained from the X-ray deepest surveys, can account for most of the XRB emission below 10 keV \citep{xue11,moretti12}. However, where the XRB spectrum peaks ($20-40$ keV), most of the XRB is still unresolved. Extrapolating to these energies, the spectrum from the deep surveys showed that an additional large population of heavily obscured, Compton-thick (CT) AGN with $N_{H}>10^{24}\,\rm{cm^{-2}}$ is required to fully reproduce the XRB spectra \citep{gilli07}. XRB synthesis models reveal that the CT-AGN may account for $10-40\,\%$ of the entire AGN population \citep[e.g.,][]{gilli07,ananna19}. \par
Finally, a complete census of these obscured accreting AGN is also needed for reconstructing the evolution of the Black Hole Accretion rate Density (BHAD) at high redshift. Generally, the shape of the BHAD resembles that of the Star Formation Rate Density (SFRD), which peaks at the ``cosmic noon'' (\textit{i.e.} $z\sim1-2$) and decreases at high-redshift. These similar trends should be linked to the AGN-galaxy co-evolution. However, most of the high-redshift measurements of the BHAD come from the deepest X-ray surveys, and those are prone to miss the most obscured AGN. In fact, at $z>3$, the BHAD estimates from the X-ray are significantly lower than what we expect from cosmological simulations \citep[\textit{e.g.}][]{volonteri16,ueda14,vito18}. This gap could be bridged if we assume the presence of a population of very obscured CT-AGN, most of which eludes our current surveys.\par
Despite their importance, obscured AGN are still not extensively studied, mostly due to the difficulty in detecting them: the large quantity of gas and dust that fuels them also has the effect of hiding them at the wavelength ranges in which we are usually selecting AGN. The X-ray radiation, originating from the hot corona in the innermost region of the AGN, is a good tracer of the AGN intrinsic emission; however, when the nucleus is obscured by column densities as large as $\sim 10^{24-25} \text{cm}^{-2}$, even hard X-rays are severely depressed \citep[e.g.,][]{comastri04,georgantopoulos13}. At high-redshifts ($z>2$), the photoelectric absorption shifts to lower energies ($<4\,\rm{keV}$, observer frame), but the detection of such obscured AGN in \textit{Chandra} and \textit{XMM-Newton} observations is still limited due to poor X-ray spectral quality \citep[e.g.,][]{kayal24}. Mid-infrared wavelengths can be effectively used as an X-ray complementary selection method, as the optical-to-X-ray absorbed radiation is re-emitted at these wavelengths after being thermally reprocessed by the obscuring torus. However, distinguishing between the AGN and the SF contributions to the total IR emission is not trivial, especially for galaxies with high SFR \citep[e.g.,][]{white25} and strong Polycyclic Aromatic Hydrocarbon (PAH) features. Future planned mid-IR cryogenic observatories (e.g., the PRobe far-Infrared Mission for Astrophysics, PRIMA,  \citealt{bradford22}), with higher sensitivity and better spectral resolution than what is currently available, will have the capabilities to both detect these sources up to very high-redshift and to distinguish the AGN and SF contribution \citep{spinoglio21, barchiesi21_spica, bisigello21a,barchiesi25}. Obscured sources can also be identified using mm-observations, as they can be sensitive up to $N_H \sim 10^{26}\,\rm{cm}^{-2}$ (e.g., \citealt{behar15,kawamuro22}). Low redshift, heavily obscured AGN with $N_H$ up to $10^{26}\,\rm{cm}^{-2}$ can also be identified using high-resolution mm-observations, which can resolve the AGN components and measure mm-luminosities that correlate with the AGN intrinsic luminosities \citep[][]{ricci23}. Finally, radio emission can be a powerful tool to identify low- and high-luminosity AGN. A major advantage of radio wavelengths is their insensitivity to obscuration, as both gas and dust extinction are negligible at typical radio frequencies \citep[e.g., $1.46\,$GHz;][]{hildebrand83} and it has been shown that the surface density of CT-AGN detected in radio surveys can be 10 times higher than in the X-rays \citep{mazzolari24}. However, radio data must be complemented by observations at other wavelengths to disentangle between AGN- and SF-related radio emission and to distinguish between obscured and unobscured AGN. \par
AGN are generally subdivided into type 1 (unobscured) and type 2 (obscured) based on the orientation of the line of sight with respect to the obscuring torus as per the unification scheme \citep{antonucci85,urry95}. With an edge-on obscuring torus, type 2 AGN can also be selected using narrow, high-ionisation emission lines, such as the \Nev, and the \Civ\ emission lines. The high-ionization potential of these lines ($\sim97\,$eV for the \nev\ and $\sim48\,$eV for the \civ) makes them an unambiguous marker of AGN activity \citep[\textit{e.g.}][]{gilli10,mignoli13,cleri22}, while the selection of type 2 AGN can be assured by excluding sources with broad spectral features. Moreover, as these lines are produced in the narrow line region (NLR), they do not suffer from nuclear extinction (i.e. extinction on pc scales due to the obscuring torus or material surrounding the SMBH), and their flux can be considered a proxy of the AGN intrinsic emission. 
Several works have shown that pairing AGN optical lines selection with X-ray data is an effective method to find obscured and CT AGN \citep[\textit{e.g.}][]{ maiolino98,cappi06,vignali06,vignali10,gilli10,mignoli13,mignoli19}. In particular, \citet[][hereafter B24]{barchiesi24} used the \nev\ lines to select obscured AGN at $z\sim1$ and to confirm that this selection method allows us to identify and characterise AGN in the obscured accretion phase of their evolution. \par
However, the characterisation of the properties of obscured AGN is not trivial, as their emission is contaminated by the host galaxy. In this case, spectral energy distribution (SED) fitting can be a powerful tool to disentangle the AGN and host-galaxy contributions, but it requires extensive multi-wavelength coverage to properly disentangle the two components. In particular, classical SED-fitting focuses mostly on the UV-to-IR wavelength range, where most of the emission of the host galaxy comes from. This is not a problem in the case of unobscured AGN, as their emission covers the same bands. However, the emission of obscured AGN peaks in the mid-IR, coinciding with the wavelength range where PAH and dust emission from the host galaxy are also present, and becomes rapidly faint outside the $1-40\,\mu$m range \citep[e.g.,][]{feltre12,hickox18}. Therefore, disentangling the host galaxy and AGN components becomes challenging. An additional difficulty comes from the large degeneracies that SED models have. With just the mid-IR emission tracing the AGN, breaking these degeneracies and properly constraining the host galaxy and AGN properties is not an easy task.\par
Luckily, in recent years, SED-fitting codes that can fit additional bands have been released. In particular, the X-ray and radio bands can be exploited to overcome these degeneracies. For the X-ray bands, the major help comes from the fact that via X-ray spectral analysis, we can correct for the source obscuration and, therefore, trace the intrinsic AGN emission. In this band, the emission linked to SF processes is usually sub-dominant with respect to the AGN one. In the radio band, both AGN activity and SF contribute to the observed emission, and distinguishing between these two processes is essential for accurate source characterisation \citep[e.g.,][]{white25}. However, one advantage is that the radio emission is not affected by dust extinction, and allows us to trace even very obscured sources. We will show that by combining optical-to-FIR, X-rays, and radio bands, we are able to overcome the degeneracy between SF and AGN emission and place better constraints on both the AGN and host-galaxy properties. \par
In this paper, we build the first sample of UV-line-selected obscured AGN with complete coverage from the X-rays to the radio band. The use of the \nev\ and \civ\ emission lines allows us to cover the entire ``cosmic noon''. Thanks to the excellent multi-wavelength coverage of the COSMOS field, we were able to perform X-ray-to-radio SED-fitting of our sources and place stronger constraints on the AGN and host-galaxy properties of our samples. We also investigate the improvement that the use of additional bands delivers in the SED-fitting. \par
The multi-wavelength observations are presented in section~\ref{sec:observations}. Section~\ref{sec:simba} introduces the \simba\ hydrodynamical simulations used as a reference for comparison with our sample. Section~\ref{sec:xray_analysis} contains the details of the X-ray spectral analysis of the \civ\ sources. In section~\ref{sec:sedfitting}, we illustrate the SED-fitting code used in this work. The results are presented and discussed in section~\ref{sec:results}. We report the conclusions in section~\ref{sec:conclusion}. Throughout this paper, we assume a Chabrier IMF \citep{chabrier03} and adopt the following cosmological parameters: H$_0 = 70\, \text{km}\, \text{s}^{-1}\, \text{Mpc}^{-1}$, $\Omega_{\text{M}} = 0.3$ and $\Omega_{\Lambda} = 0.7$ \citep{spergel03}. The uncertainties associated with the median value are computed from the $16^{\rm{th}}-84^{\rm{th}}$ percentiles, unless stated otherwise. For X-ray spectral fitting derived properties, the uncertainties are computed at $90\%$ confidence level. Finally, unless stated otherwise, with

\section{Observations}\label{sec:observations}

\subsection{Sample selection}
Our sample comprises 184 obscured AGN, 94 are part of the \nev\ type 2 AGN sample of \citet[][hereafter M13]{mignoli13}, 90 come from the \civ\ type 2 AGN sample \citet[][hereafter M19]{mignoli19}. We will refer to the two samples individually as the \nev\ sample and the \civ\ sample, respectively.\par
The \nev\ sample was drawn from the zCOSMOS-Bright spectroscopic survey \citep{lilly07, lilly09}, which provided the $5500 - 9700$\AA\ spectra of $\sim 20000$ objects in the COSMOS \citep{scoville07} field. M13 focused on the sources in the $\sim 0.65-1.20$ redshift range (the redshift range assured that both the \nev3346\AA\ and the \nev3426\AA\ emission line fall within the spectral coverage), with \nev\ emission line detection, and only narrow features in their spectra (i.e. to exclude type 1 AGN). We refer to M13 for an in-depth analysis of the sample selection. \par
The \civ\ sample was drawn from the zCOSMOS-deep survey. Briefly, the zCOSMOS-deep survey includes 9523 spectroscopically observed objects, 80\% of them with available redshift. By limiting the sample to the $1.45-3.05$ redshift range (4391 galaxies), it is guaranteed that the \civ\ line fell within the observed wavelength range. Their VIMOS spectra were visually inspected to identify \civ\ emitters. Finally, all the sources with a \civ\ emission peak five times higher than the nearby (i.e., in a 50\AA\ windows around the \civ\ line) continuum RMS were classified as \civ\ emitters, resulting in a sample of 192 AGN. By adopting a $2000\,\kms$ FWHM threshold, 102 were classified as type 1 AGN and 90 as type 2 AGN, which were included in the final sample for this work. The adopted threshold between type 1 and type 2 refers to the measured FWHM plus its associated uncertainty, and its choice is supported by the fact that type 1 AGN were already identified in the previous steps by the presence of any broad line (not just the \civ\ line) in their spectra. We refer to M19 for an in-depth description of the sample selection. \par
When taking into account that the two samples have been built with different thresholds for the detection of the emission lines (signal-to-noise ratio of 2.5 and 5 for the \nev\ and \civ, respectively), trace different redshift ranges, and that the \nev\ emission line is usually vary faint, while the \civ\ line is one of the brightest in an AGN spectra, we expect the \civ\ sample to possibly include intrinsically weaker emitter than the \nev\ sample. However, we also caution that the \civ\ line is quite sensitive to gas metallicity \citep{mignoli13,dors14,dors25},  while the \nev\ emission can be significantly attenuated by absorption in the host-galaxy ISM.\par
\begin{figure}
  \centering
  \resizebox{0.8\hsize}{!}{\includegraphics{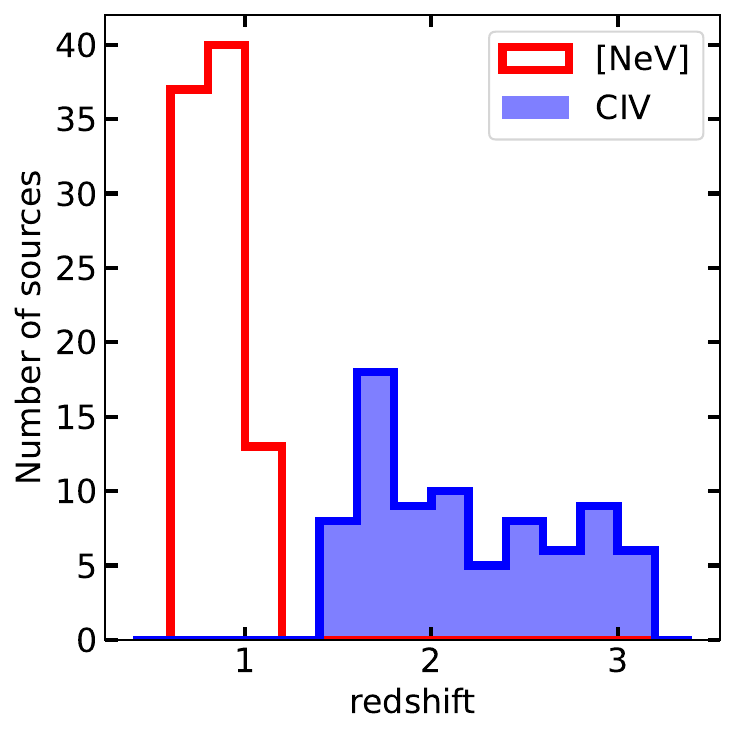}}
  \caption{Redshift distribution of the \civ- (\textit{blue}) and \nev- (\textit{red}) selected AGN. The redshift distribution is primarily determined by the requirement that the \Nev\ and \Civ\ emission lines fall within the wavelength coverage of VIMOS.}
  \label{fig:redshift_hist}
\end{figure}

In Fig.~\ref{fig:redshift_hist}, we show the redshift distribution of the \nev\ and \civ\ samples. 

\subsection{Multiwavelength data in the COSMOS field}

\subsubsection{X-ray}\label{sec:xray}
All 184 Type 2 AGN falls in the \textit{Chandra}-COSMOS Legacy mosaic, which is composed of data from the C-COSMOS survey \citep[the central $\sim0.9\,$deg$^2$;][]{elvis09} and from the COSMOS Legacy survey \citep[covering the external $\sim 1.7$ deg$^2$ with a similar depth of the C-COSMOS survey;][]{civano16}. The whole mosaic covered $\sim 2.2$ deg$^2$ with a total exposure time of $\sim 4.6$ Ms. The fitting of the X-ray spectra of the \nev\ AGN has been presented in B24, and we will refer to their values: of the 94 \nev\ AGN, 36 are X-ray detected, with median intrinsic $2-10$ keV rest-frame luminosity of $\log{(L_{\rm{2-10keV}}^{\rm{\nev}}/\rm{erg\,s^{-1}})}=43.6_{-0.4}^{+0.6}$, and obscuration of $\log{(N_{\rm{H}}^{\rm{\nev}}/\rm{cm^{-2}})}=22.6^{+0.7}_{-0.9}$. For the X-ray undetected sources, by exploiting the upper limit on their X-ray flux and the \citet{gilli10} correlation between the $N_H$ and the $X/$\nev\ ratio, B24 obtained a lower limit on their obscuration, with $\sim90\%$ having $\nh \geq 10^{23}\rm{cm^{-2}}$ and $\sim30\%$ being candidate CT-AGN ($N_{\rm{H}}>10^{24}\,\rm{cm^{-2}}$).\par
For the \civ\ sample, we performed a similar analysis as B24, which is presented in section~\ref{sec:xray_analysis}.

\subsubsection{Optical - Near Infrared}
We collected the optical-NIR photometry of our sources from the COSMOS 2020 catalogue \citep{cosmos2020} (improved version of the previous COSMOS 2015 catalogue \citealt{laigle16}), which contains photometry in 44 bands (from 1526\AA\ to 8$\mu$m) for $\sim 1.7$ million objects in the 2 deg$^2$ COSMOS field, along with matches with X-ray, near-ultraviolet (near-UV), and radio data. The COSMOS catalogue includes far-UV and near-UV bands from the Galaxy Evolution Explorer \citep[GALEX,][]{zamojski07}, \textit{u}$^*$-band data from the Canada-France-Hawaii Telescope Legacy Survey \citep[CFHTLS,][]{sawicki19}, \textit{g, r, i, z, y} band from the Hyper Suprime Cam \citep[HSC,][]{aihara19}, \textit{Y, J, H, K}$_S$ band from the UltraVISTA Survey \citep{mccracken12,moneti19}, and 3.6 and $4.5\,\mu$m from the \textit{Spitzer} Infrared Advanced Camera \citep[IRAC,][]{ashby13,steinhardt14,ashby15,ashby2018}.\par
We cross-matched our sources with the COSMOS2020 Classic catalogue on the basis of the source positions, collected the 2" aperture fluxes from the catalogue, and, via the COSMOS2020 pipeline, corrected those for the Milky Way extinction, aperture, and photometric offset.\par

\subsubsection{Mid- and Far-Infrared}
The FIR photometry of our sources comes from the XID+ deblended catalogue \citep{wang24} and comprises \textit{Spitzer} MIPS $24\,\mu$m, \textit{Herschel} PACS $100\,\&\,160\,\mu$m, and \textit{Herschel} SPIRE $250\textnormal{,}\, 350\, \&\, 500\,\mu$m. \citet{wang24}  used a probabilistic and progressive deblending algorithm to deblend point-like sources in the COSMOS field down to the $1 \sigma$ noise. In particular, starting from a prior list of galaxies from the COSMOS2020 catalogue, and from VLA and MeerKAT surveys, and exploiting SED fitting modelling, they were able to predict flux densities at MIPS $24\,\mu$m. The deblended data at $24\,\mu$m, were then used as priors to deblend the PACS $100\mu$m, and subsequently  PACS $160\,\mu$m and SPIRE at $250\textnormal{,}\, 350\, \&\, 500\,\mu$m. We refer to \citet{wang24} for further details.\par
We cross-matched our sources with the XID+ catalogue based on the source position. We also visually inspected each source to confirm the reliability of the matching and flagged the fluxes that may be potentially contaminated by neighbouring sources. Of our sample, $74\%$ of the sources are detected at $24\mu\rm{m}$ with \textit{Spitzer MIPS}, $64\%$ are detected with \textit{Herschel PACS} either at $100\mu\rm{m}$ or $160\mu\rm{m}$, and $48\%$ have detection at any of the three \textit{Herschel SPIRE} bands.

\par

\subsubsection{Radio}\label{sec:data_radio}
For this work, we used radio data from the MeerKAT International GHz Tiered Extragalactic Exploration Survey (MIGHTEE) \citep{jarvis16} and the VLA-COSMOS 3GHz Large Project \citep{smolcic17a,smolcic17b}.\par
The VLA data consist of 384hr observations in \text{S}-band ($2-4\,$GHz), with a resolution of $0.78''$ and a sensitivity of $2.3\,\mu\text{Jy}\,\text{beam}^{-1}$. The MeerKAT data are from the Data Release 1 of the MIGHTEE continuum survey, fully described in \citet{hale25}. We also used the MIGHTEE multi-wavelength counterpart association of \citet[][hereafter W24]{whittam24} for the comparison of our source properties with the average MIGHTEE-detected galaxies. Briefly, the MIGHTEE survey covers the COSMOS, XMM Large-Scale Structure (XMM-LSS), Chandra Deep Field South (CDFS), and European Large Area ISO Survey (ELAIS) S1 ﬁelds with the MeerKAT \textit{L}-band receiver ($856-1711\,$MHz). The MIGHTEE observations of the COSMOS field cover $4.2\,\text{deg}^2$ for $139.6\,$hr on-target. We note that the counterpart association of W24, which is based on the Early Science data \citep{heywood22}, covers only the central $0.86\,\text{deg}^2$. We used both the ``high-resolution'' image, processed using a Briggs robust value of $-1.2$ for a resolution of $\sim5.2''$ and a median measured central rms sensitivity of $2.4\,\mu\text{Jy}\,\text{beam}^{-1}$, and the ``low-resolution'' one, with a Briggs robust of $0$, resolution of $8.9''$, and rms of $2.1\,\mu\text{Jy}\,\text{beam}^{-1}$. As described in \citet{hale25}, the source extraction was performed using the Python Blob Detection and Source Finder \citep[\texttt{PYBDSF},][]{pybdsf}.\par
We cross-matched our sources with both the VLA and MIGHTEE catalogues, based on their position on the sky with matching radii of $1.5''$ and $4''$, respectively. We also visually checked the MIGHTEE, VLA, SUBARU, and UltraVISTA images to confirm the radio association and to flag the MIGHTEE sources for which a secure association was not possible (mostly due to having more than one source within their beam). 
We found that out of our 184 sources, 107 have a possible counterpart in MIGHTEE, 66 are from the \nev\ sample and 41 from the \civ\ one. 79 sources have a secure counterpart association for both the ``high-'' and ``low resolution'' images, 19 (1) only in the ``high (low) resolution'' image, and 8 do not have a secure counterpart identification. The higher number of secure counterparts in the high resolution (but lower sensitivity) catalogue is driven by the radio emission of these sources having multiple possible associations that we are able to resolve in the higher-resolution image and not in the lower one. We found that 81 sources have a secure detection in the VLA data, of which 76 are also detected in MIGHTEE. For the sources without radio counterparts in MIGHTEE (VLA) data, we used the average RMS at the source position to compute their 5$\sigma$ flux density upper limits. We report in Table~\ref{tab:detections_radio} the number of \nev\ and \civ\ with ``secure'' detection in the MIGHTEE and VLA surveys. Table~\ref{tab:detections} shows the number of detected and undetected sources in the radio and X-ray bands. 

\begin{table}
\caption{Number of the \nev\ and \civ\ sources in the MIGHTEE survey at $\sim1.24\,\rm{GHz}$ and in the VLA-COSMOS at  $3\,\rm{GHz}$. \textit{detected} refers to a ``secure'' detection, i.e. detection and secure counterpart association. \textit{undetected} means either a non-detection or that it was not possible to unequivocally identify its counterpart. }
\label{tab:detections_radio}
\centering
\begin{tabular}{ccccc}
\hline 
 & MIGHTEE detected  & MIGHTEE undetected  \\
\hline
 VLA detected & 76 & 5 \\
  VLA undetected & 31  & 72 \\
\end{tabular}
\end{table}

\begin{table}
\caption{Number of detected \nev\ and \civ\ sources in the Chandra COSMOS Legacy survey and in the radio band, i.e., in the MIGHTEE survey at $\sim1.24\,\rm{GHz}$ or in the VLA-COSMOS at 3GHz. \textit{detected} refers to a ``secure'' detection, i.e. detection and secure counterpart association. \textit{undetected} means either a non-detection or that it was not possible to unequivocally identify its counterpart. }
\label{tab:detections}
\centering
\begin{tabular}{ccccccc}
\hline 
 & & Radio detected  & Radio undetected \\
\hline
  \multirow{2}{*}{\nev} &  X-ray detected & 31 & 5 \\
 & X-ray undetected & 37 & 21 \\[3pt]
  \multirow{2}{*}{\civ} & \ X-ray detected & 30 & 22 \\
 & X-ray undetected & 14 & 24 \\
\end{tabular}
\end{table}

\section{Hydrodynamical simulations}\label{sec:simba}
To study the evolution of our \nev\ and \civ\ sources and to compare their properties with the general population of galaxies, we exploited the results of the \simba\ hydrodynamical simulations. \simba\ \citep{dave19} is a cosmological suite of simulations designed to study the co-evolution of galaxies, BH, and the intergalactic medium. We refer to \citet{dave19} for a comprehensive description. Briefly, we employed the $100\,h^{-1}\,\rm Mpc$ simulation box, which evolves $1024^3$ dark matter particles and an equal number of gas elements from $z = 249$ to $z = 0$, using the \textsc{Gizmo} code in its Meshless Finite Mass (MFM) configuration \citep{hopkins15,hopkins17}. The simulations incorporate radiative cooling and photoionisation heating via the \textsc{Grackle-3.1} library \citep{smith17}, which includes metal cooling and non-equilibrium chemistry for primordial elements. Star formation is based on local H$_2$ column density and metallicity, while galaxy properties, such as stellar mass, are identified using an on-the-fly friends-of-friends finder applied to stars and dense gas.\par

Black holes are seeded and grown during the simulation, with the accretion energy used to drive feedback and to quench the SF. The accretion is implemented with two different modes: a torque-limited mode, which models cold gas inflows ($T < 10^5\,\rm K$) driven by disk instabilities and capped at three times the Eddington limit, and a Bondi accretion mode for hot gas ($T > 10^5\,\rm K$) under spherical symmetry, capped at the Eddington rate. A constant radiative efficiency of $\eta = 0.1$ is assumed in both cases. AGN feedback is modelled via two subgrid prescriptions: a ``radiative'' mode active at high Eddington ratios ($\eddrat \gtrsim$ a few percent), which drives molecular and warm ionised outflows, and a ``jet'' mode at low Eddington ratios, where the AGN mostly drive hot gas in collimated jets at high velocities. \par

\section{X-ray spectral analysis}\label{sec:xray_analysis}
For the \civ\ AGN, we performed the X-ray data reduction and fitting following the same procedure as in B24.\par
Briefly, the source spectra were extracted using the \textit{CIAO} \citep{ciao} tool \texttt{specextract} from circular regions of radius $r_{90}$ (i.e., the radius that contains 90\% of the PSF in the 0.5 - 7 keV observed-frame band), and the background spectra from annuli centred on the source position. The spectra from different observations were combined in a single one via the \textit{CIAO} tool \texttt{combine\_spectra}. Of the 90 \civ\ AGN, 52 had an X-ray counterpart, with a median (and $16^{\rm{th}}-84^{\rm{th}}$ percentiles) number of net-counts of $48_{-22}^{+36}$.  Two sources had fewer than 20 counts, and only one source had more than 200 counts. Due to the low number of net-photons, we used unbinned data and \textit{C}-statistic \citep{cash79} for all the sources. The higher X-ray detection fraction of the \civ\ AG with respect to the \nev\ ones is due to the combination of two factors: at higher redshifts, we are less affected by the obscuration of the source (i.e., we are able to detect more obscured AGN as we are sampling higher rest-frame energy); secondly, the effective volume that we are sampling with the \civ\ selection is $\sim 4.6$ times higher than the one of the \nev\ sources.\par
To fit the spectra, we choose two simple phenomenological models. The first was composed of a power-law corrected for the Galactic absorption to obtain the spectral slope. The second one had a fixed $\Gamma=1.8$ power-law (representing the AGN intrinsic emission) and an additional absorption component to model the AGN obscuration. The choice of simple phenomenological models was driven by the low number of photons, which did not allow for physically motivated but more complex modelling. \par
We visually inspected all the spectral fittings and added more complex components if the fit was not considered satisfactory (if the number of counts allowed it) and if the additional component had a significance $>3\,\sigma$. As we used cash statistics, we were not able to exploit the F-test to compute the additional component significance; instead, we analysed the 2D contours of the free parameters of the component. If we were able to ``close'' the $3\,\sigma$ contours, we considered the component significant. For example, for source zCOSMOS ID: 403380, we obtained a significantly better fit with a partial covering component with a covering fraction of $f=0.95_{-0.08}^{+0.03}$.\par

Finally, 22 sources had significant contributions of the background in their spectra or background residual in some bins of their spectra (most of them have less than 40 net-counts and all have $N_H>10^{23}\,\rm{cm^{-2}}$. For those, we chose to perform an additional run of X-ray spectral fitting, modelling the background instead of just subtracting it. The background was modelled with a phenomenological model, following \citet{fiore12}; briefly, it consists of two power-law components (to reproduce the continuum), three narrow Gaussian components (to model the emission lines at 1.48, 1.74, and 2.16$\,$keV), one broad Gaussian component (to reproduce the broad bump between 1 and 2$\,$keV), and a thermal component. We fitted the background of each source separately, and not all the components were present in all the spectra. Once we found a good fit for the background, we rescaled its normalisation with the ratio between the extraction area of the source and the area of the background and used it alongside the source model to fit the AGN spectrum. For all these sources, we then computed the intrinsic luminosity and amount of obscuration, assuming the same AGN model as the background-subtracted sources. \par
The median spectral index, intrinsic luminosity, and obscuration of the \civ\ (and \nev) AGN are reported in Table \ref{tab:civ_xray_prop}. Based on their $\rm{N_{\rm{H}}}$, almost all the X-ray detected sources (48) are obscured ($N_H\geq10^{22}\,\rm{cm^{-2}}$), 37 (71$\,$\%) are extremely obscured ($N_H>10^{23}\,\rm{cm^{-2}}$), and 2 (4$\,$\%) are CT-AGN ($N_H\geq10^{24}\,\rm{cm^{-2}}$).\par
\begin{figure}
  \centering
  \resizebox{0.7\hsize}{!}{\includegraphics{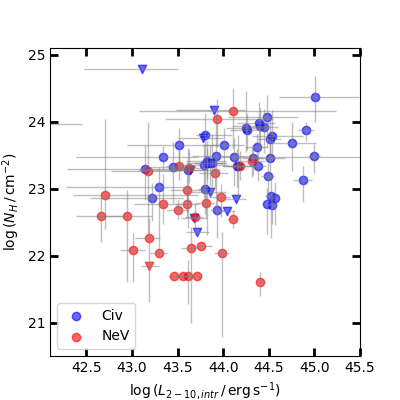}}
  \caption{Obscuration as a function of the intrinsic $2-10\,$keV luminosity for X-ray detected sources in the \civ\ (\textit{blue}) and \nev\ samples (\textit{red}). The \textit{grey lines} are the associated $90\%$ uncertainties. The \textit{triangles} denote upper limits on the amount of obscuration due to the difficulties in properly constraining it in the case of spectra with a low number of photon counts.}
  \label{fig:civ_lx_nh}
\end{figure}

From Fig.~\ref{fig:civ_lx_nh}, it is easy to see that the \civ-selected AGN (at least those with X-ray detection) are more luminous and more obscured than the \nev\ ones. The upper limits on the AGN obscuration is a consequence of the low number of X-ray photons: even with a simple phenomenological model, for some sources we were not able to properly constrain the $N_H$ at the $90\%$ confidence level. Although the Malmquist bias plays a role in the average higher luminosity of the \civ\ sample, it is not enough by itself to justify the difference. If we cut both \nev\ and \civ\ samples to the luminosity limit of the survey at $z=3$, the \civ\ sources are still more luminous than the \nev\ sources, suggesting that this difference may be intrinsic and possibly due to the use of two different lines for the selection. In fact, as it is discussed in section~\ref{sec:disc_diff_pop}, the two lines select different AGN populations that will evolve into galaxies with different properties.

\begin{table}
\caption{X-ray spectral properties (median and $16^{\rm{th}}-84^{\rm{th}}$ percentiles) of the \civ\ and \nev\ samples. $\Gamma$ is the photon index obtained from a simple power-law (corrected for the Galactic absorption) model. In this model, an obscured source has a lower (flatter) photon index than an unobscured source. $L_{\rm{2-10,intr}}$ and $N_{\rm{H}}$ are the intrinsic (i.e., absorption corrected) $2-10\,$keV luminosity and the amount of obscuration, obtained from a power-law model (corrected for the Galactic absorption) with fixed photon index $\Gamma=1.8$ plus an absorption component.}
\label{tab:civ_xray_prop}
\centering
\begin{tabular}{ccc}
\hline 
 & \civ\ & \nev\ \\
\hline
$z$ & $2.16_{-0.50}^{+0.56}$ & $0.88_{-0.17}^{+0.10}$ \\[3pt]
$N_{\rm{net-counts}}$ & $48_{-22}^{+36}$ & $83_{-63}^{+129}$\\[3pt]
$\Gamma$ & $0.6_{-0.6}^{+0.7}$ & $1.1_{-1.1}^{+0.6}$\\[3pt]
$\log\,(\,L_{\rm{2-10,intr}}\,/\,\rm{erg\,s^{-1}}\,)$ & $44.1_{-0.5}^{+0.4}$ & $43.6_{-0.6}^{+0.4}$\\[3pt]
$\log\,(\,N_{\rm{H}}\,/\,\rm{cm^{-2}}\,)$& $23.4_{-0.6}^{+0.4}$ & $22.6_{-0.9}^{+0.7}$\\[3pt]
\hline
\end{tabular}
\end{table}

\section{SED fitting}\label{sec:sedfitting}
There are several codes available to perform SED-fitting; we choose the Code Investigating GALaxy Emission (\texttt{CIGALE}), which has been developed to study the evolution of galaxies by comparing modelled galaxy spectral energy distributions to observed ones. The code was presented in \citet{burgarella05,boquien19} and improved with the addition of the X-ray and radio modules in \citet{yang20,yang22}.\par
In \texttt{CIGALE}, the SED models are built through a series of ``modules'' defined by the user. This architecture allows us to easily customise and tailor the code for our specific use. Unfortunately, the need to model the emission in both the X-ray and radio parts of the spectrum significantly increases the computational cost of the SED-fitting, in terms of both time and resource allocation. For this reason, we tried multiple runs of SED-fitting, starting from a sparse sampling of the components and iteratively increasing the sampling at each run. In every run, we checked which parameters were the most effective in obtaining a better fit, and which were instead heavily degenerate. In this regard, a huge help came from the comparison with mock SED: for each best-fit model, we simulated mock observations adding random noise based on the filter uncertainty, re-run the SED-fitting, and compare the ``real'' properties with those obtained from the mocks. If a parameter showed consistent values in the ``real'' and mock results (i.e., if the two agreed within their uncertainties, and those uncertainties were smaller than the full allowed range of that parameter), we considered it well constrained. Conversely, if the ``real'' and mock values differed significantly, typically with large uncertainties, this indicated that the parameter could not be reliably constrained through the SED fitting of our observations. The latter was the case for parameters that produce subtle differences in the SED - which would require narrower filters or even spectroscopy to be properly studied, e.g., the torus opening angle, the torus outer-to-inner radius ratio, the galaxy nebular emission - or combinations of parameters that suffer from heavy degeneracy that cannot be resolved with our coverage (e.g., the torus optical depth and its density profile).\par
Finally, after fitting all the sources with a ``simple'' model ( with more than $10^9$ parameter combinations), we used more complex or different parameter combinations only for the subsample that we could not properly fit.\par
For the SED-fitting, we use a delayed star formation history (SFH) with an optional exponential burst, the \citet{bruzual03} population synthesis model with a \citet{chabrier03} initial mass function, nebular emission lines, a \citet{calzetti00} dust attenuation law, and the \citet{draine14} dust models. The AGN emission is implemented via the SKIRTOR models \citep{stalevski12,stalevski16}, in which the torus is modelled as a clumpy two-phase medium \citep[we refer to][for further details]{yang20}. As we know from the sample selection that our sources are type 2 AGN, we chose only AGN inclination for which our line of sight would intercept the torus, i.e. $i \geq 90^{\circ} - oa$, where $oa$ is the angle measured between the equatorial plane and the ``edge'' of the torus. In our analysis, we also found that a moderate sampling of the AGN extinction in polar direction is needed to obtain better fitting, as this parameter influences the shape of the IR AGN emission near its peak, i.e. where our \textit{Herschel} filters lie. 
We make use of the X-ray module, which takes into account the emission in the X-ray from the AGN and from the X-ray binaries of the host galaxy. Finally, we also used the radio module, which has two components representing the emission of the AGN and the one related to SF. In Table~\ref{tab:civ_cigale}, we report the modules and parameters used for the SED-fitting.\par
As a sanity check, we tried some runs of SED-fitting using the FRITZ2006 torus models \citep{fritz06,feltre12}. Similarly to other works on high-redshift galaxies \citep[e.g.][]{barchiesi23,barchiesi24}, the mid-, far-IR available coverage does not allow us to distinguish between the two models. In particular, the SEDs seem equally well fitted by both models, and the main AGN parameters do not show significant differences.\par
Finally, we tried the LOPEZ24 module \citep{lopez24}, which is an X-ray module tailored for fitting low-luminosity AGN (LLAGN) and AGN with an Advection Dominated Accretion Flow \citep[ADAF,][]{xie12} structure rather than a standard accretion disk \citep[e.g.,][]{shakurasunyaev73}. The other main difference of between the SKIRTOR and LOPEZ24 modules is that, in the latter, the X-ray emission is linked to the intrinsic AGN component via the $L_X-L_{12\mu\text{m}}$ \citep[e.g.][]{gandhi09}, instead that via the $L_{2\text{keV}}-L_{2500\AA}$.\par

\begin{table*}
\caption{Parameter space for the \texttt{CIGALE} SED fitting. For each parameter, \rm{N$_{\text{sample}}$} values are simulated in the \textit{values} range. For each module, we report only the parameters that we changed with respect to the default values. We refer to \citet{boquien19,yang22} for the complete list.}
\label{tab:civ_cigale}
\centering
\resizebox{1.8\columnwidth}{!}{\begin{tabular}{cccc}
 & N$_{\text{sample}}$ & values & description \\
\hline \hline
SFH & & sfhdelayed & delayed SFH with optional exponential burst\\
$\tau_{\text{main}}$ & 8 & $50,100,200,500,700,1000,2000,5000$ & e-folding time of the main stellar population model in Myr.\\
age$_{\text{main}}$& 10 & $200,500,700,1000,2000,3000,4000,5000,7000,9000$& Age of the main stellar population in the galaxy in Myr.\\
$\tau_{\text{burst}}$ & 1 & 1000 & e-folding time of the late starburst population model in Myr.\\
age$_{\text{burst}}$ & 4 & $20,50,70,100$ & Age of the late burst in Myr.\\
$f_{\text{burst}}$ & 4 & $0.0, 0.001, 0.01, 0.1$ & Mass fraction of the late burst population.\\
\hline
IMF& & Chabrier & Initial Mass Function\\
$Z$ & 1 & $0.02$&Metallicity\\
\hline
Z$_{\text{gas}}$& 1 & $0.02$&Nebular component: gas metallicity\\
\hline
dust attenuation& & Calzetti 2000\\
E(B-V)$_{\text{lines}}$ & 7& $0.05, 0.1, 0.2, 0.3, 0.4, 0.5, 0.6$ & Color excess of the nebular line light.\\
E(B-V) factor & 3 & $0.25, 0.44, 0.75$ & $\rm{E(B-V)}_{\text{continuum}} / \rm{E(B-V)}_{\text{lines}}$ factor.\\
\hline
Dust emission& & Draine+2014&\\
 $\alpha$& 3 & $ 1.5, 2.0, 2.5$ & Powerlaw slope $dU/dM \propto U^{\alpha}$. \\
\hline
AGN& & Skirtor16\\
$\tau_{9.7\mu\text{m}}$& 2 & $7,11$& Average edge-on optical depth at 9.7$\,\mu$m.\\
$i$ & 4 & 60\textdegree, 70\textdegree, 80\textdegree,90\textdegree & Viewing angle (w.r.t. the AGN axis).\\
f$_{\text{AGN}}$& 11 & $0,0.1,0.2,0.3,0.4,0.5,0.6,0.7,0.8,0.9,0.99$ & $1-1000\,\mu$m AGN fraction.\\
$E_{B-V}$ & 8 & $0,0.03,0.06,0.9,0.12,0.15,0.2,0.4$ & $E_{B-V}$ for the extinction in the polar direction in magnitudes \\
\hline
X-ray & \\
$\alpha_{\text{ox}}$ & 5 & $-1.9,-1.7,-1.5,-1.3,-1.1$ & $\alpha_{\text{ox}} = 0.3838\times\log(L_{\nu,2\text{keV}}/L_{\nu,2500\AA})$\\
\hline
radio & \\
q$_{\text{IR}}$ & 2 & $2.4,2.5,2.6,2.7 $ & FIR/radio correlation coefficient for SF.\\
$R_O$ & 8 & $0.01, 0.02, 0.05, 0.1, 0.2, 0.5, 1, 2, 5, 10, 20$ & AGN optical radio-loudness $R_O=L_{\nu,5\text{GHz}}^{AGN}/L_{\nu,2500\AA}^{AGN,\text{intr}}$ at $i=30$\textdegree.\\
\hline
\end{tabular}}
\end{table*}\par
\par
As for the current version, \texttt{CIGALE} requires that the input X-ray fluxes are intrinsic. This means that the fluxes by definition should already be corrected for the instrumental response and for any obscuration of the AGN itself. To this goal, we used the intrinsic $0.5-2\,$ and $2-10\,$keV (observer-frame) flux that we obtained from the X-ray spectral analysis (Section~\ref{sec:xray}). Finally, we used box-shaped filters for the two filter responses (a fair assumption as we have already corrected for the instrumental response).\par
Finally, regarding the radio fluxes, we use two box-shaped filter responses for the VLA and MeerKAT observations. For MeerKAT photometry, we centred the filter response at $1.24\,$GHz, the median effective frequency at our source positions. The use of the median effective frequency for all the sources, instead of using a different frequency for each source, was simply driven by the small difference in effective frequency of our sources (median variation of $\sim0.02\,$GHz) and computationally convenience: using a different frequency means that we had to build a response function for each source and use a slightly different filter for each source. The relative flux error due to our assumption is $<2\%$, less than the median uncertainties of the MIGHTEE fluxes (i.e. $\sim 10\%$).\par
For the SED fitting, we used the $1\sigma$ uncertainty associated with the fluxes in their catalogues and an additional uncertainty of $10\%$ of the flux added in quadrature, as common practice in CIGALE. 

\subsection{Goodness of fit}
Of our entire sample of 184 sources, we are not able to obtain a good fit of the SED for 12 of them, either due to a reduced $\chi^2>10$ or due to clearly suboptimal fits with FIR residuals larger than several times the associated uncertainties. Except for a few peculiar cases, the latter can be mostly attributed to some mismatch between the optical and the IR part of the spectrum, with the latter severely over- or under-estimated with respect to measured fluxes. Excluding cases where contamination from IR-unresolved sources may affect the photometry—particularly for the \civ\ sources, which tend to have lower fluxes and smaller sizes—one contributing factor to the observed discrepancies is the uneven distribution of photometric coverage and associated uncertainties. Firstly, the optical regime of our SED has significantly better coverage than the IR. Secondly, the uncertainties in the IR are systematically larger than those in the optical. These two factors lead to the optical part of the spectrum having a higher weight than the IR one when minimising the $\chi^2$.\par
For the sources for which we are able to obtain a good SED-fitting, we have a median (and $16^{\text{th}}-84^{\text{th}}$ percentile uncertainties) reduced chi-squared of $\Tilde{\chi}^{2}=1.3_{-0.5}^{+1.0}$, with no significant difference between the \nev\ and \civ\ sample. 30 sources are better fitted with an additional recent ($<50\,$Myr old) burst to their SFH, which contributes $\sim 0.1-1\%$ to the total stellar mass of the galaxy. For 20 of these sources, the improvement in the fit is significant even taking into account the presence of the two additional free parameters ($\tau_{\text{burst}}$ and $\text{age}_{\text{burst}}$). The difference of the Akaike information criteria \citep{akaike74} are $\Delta \text{AIC}$<-4, corresponding to the bursty models being $>50$ times more probable than the non-bursty ones \citep{Burnham02,yang22}. Finally, 8 of the 30 sources showed only a small improvement to their $\chi^2$, but their best-fits were visually significantly better: with the bursty model being able to properly reproduce either the FIR or the X-ray part of the spectrum. For all the sources, we did not find any significant difference with respect to the whole population regarding their X-ray or radio properties. 
Finally, except for one source (zCOSMOS ID 841734), we obtained significantly worse fits (both visually and $\chi^2$ wise) when using the LOPEZ24 X-ray module, indicating that our AGN are more likely to have classical ``optically thick - geometrically thin'' disks rather than ADAF ones. Regarding zCOSMOS ID 841734, we obtained significantly better fit and lower reduced $\chi^2$ ($\Tilde{\chi}^{2}=5.6$ vs $ \Tilde{\chi}^{2}=6.2$, with a $\Delta \text{AIC}=-39$) using the LOPEZ24 module, with a bolometric luminosity of $\log{(L_{\rm{bolo}}/\rm{erg\,s^{-1}})}=45.8\pm0.1$. The relatively high bolometric luminosity makes this source an interesting puzzle, as it does not fall in the classical classification of LLAGN \citep[$L_{\rm{bolo}}<10^{44}\,\rm{erg\,s^{-1}}$, e.g. ][]{peterson97}, and ADAF structures are expected at very low accretion rate \citep[$\lambda_{\rm{Edd}}<10^{-3}$][]{yuan14} - although ADAF-like models (e.g., Luminous Hot Accretion Flow, LHAF) have been theorised for higher luminosity AGN \citep{yuan03,yuan04,xie16}. \par

\subsection{SED fitting parameter reliability}
\begin{figure}
	\includegraphics[width=\columnwidth]{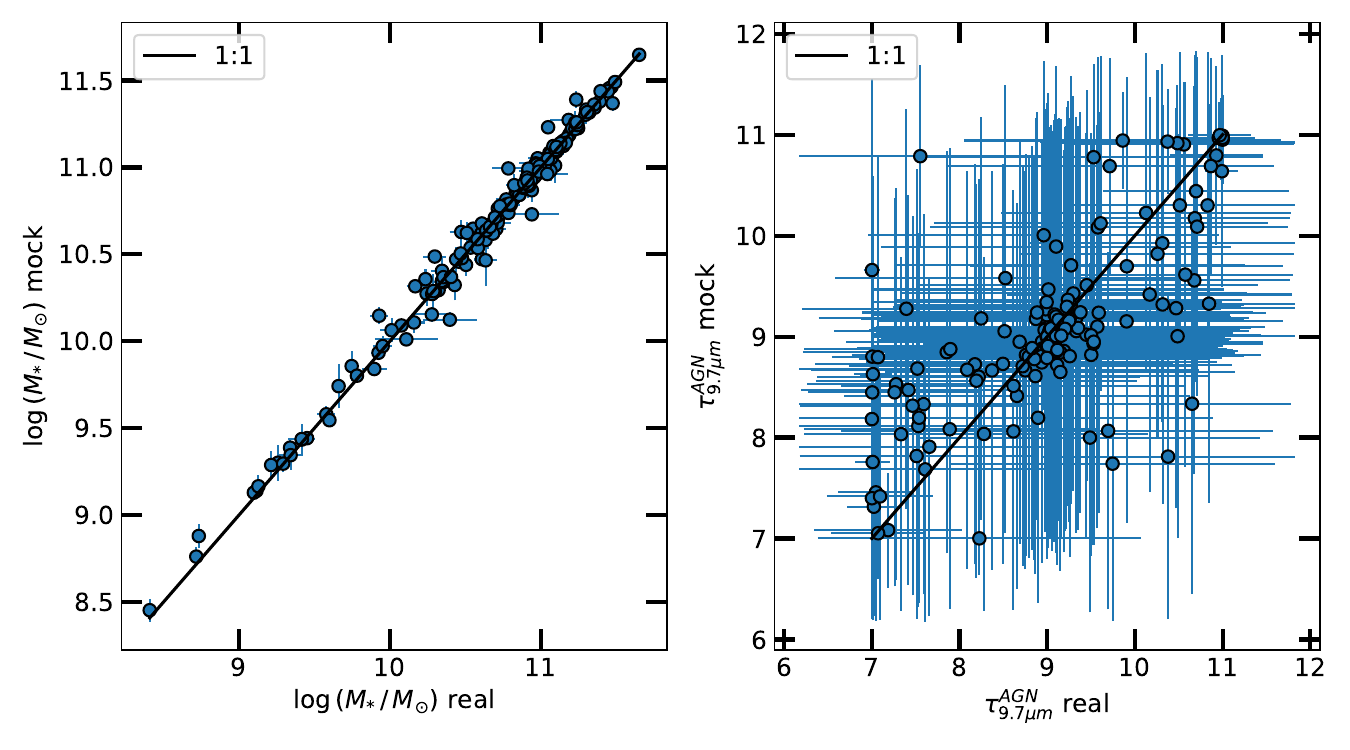}
    \caption{Comparison of the stellar mass (\textit{left}) and AGN torus optical depth (\textit{right}) obtained from SED fitting for the real and mock observations.  While we can robustly estimate the host-galaxy stellar mass, with the available data, we are not able to put proper constraints on the AGN optical depth.} 
    \label{fig:sed_mock}
\end{figure}

By comparing our results with those from the mock simulations, also taking into account the uncertainties, we can confirm if the AGN and galaxy parameters obtained from the SED fitting are nicely constrained or heavily degenerate. As an example, we show in Fig.~\ref{fig:sed_mock} the comparison of the ``real'' and mock values of the stellar mass and of the AGN $9.7\,\mu$m optical depth. It is easy to see that the stellar mass is nicely constrained, with small uncertainties and best-fit values that are not really affected by the observational uncertainties. The case is different for the $9.7\,\mu$m optical depth, which cannot be constrained with our current photometric coverage. Minor flux variations lead to large uncertainties in the best-fit $9.7\,\mu$m values. This is primarily because different $9.7\,\mu$m values induce only subtle changes in the overall SED, which would require IR spectroscopy to be measurable \citep[e.g.,][]{feltre12}. 

Regarding the main AGN and host-galaxy properties, we find that the stellar mass, SFR, and AGN bolometric luminosity are nicely constrained by our fitting. We note, however, that the uncertainties on the AGN bolometric luminosities can be larger than one magnitude for the sources with luminosity $L_{\text{bolo}}<10^{45}\,\text{erg}\,\text{s}^{-1}$. The AGN fraction, optical radio loudness $R_O$, $\alpha_{OX}$, AGN extinction in polar direction $E^{AGN}(B-V)$, and host-galaxy dust attenuation are constrained too, but with larger uncertainties (see Table~\ref{tab:civ_cigale}). Finally, we report that we were not able to put proper constraints on the AGN optical depth $\tau_{9.7\mu m}$, opening-angle, outer-to-inner radius ratio, torus radial- and polar density gradient, host-galaxy metallicity, and nebular line properties.

\section{Results}\label{sec:results}

\subsection{Importance of Radio band in SED-fitting}\label{sec:importance_radio_sed}
To investigate the impact of radio band coverage on SED fitting, we performed additional runs of SED fitting, using the same models and parameters but excluding the radio band. We compared the results and the goodness of fit obtained with and without the inclusion of the radio data. Our main finding is that the effects of including the radio band depend on the quality of the multi-wavelength coverage. We present three representative scenarios, the SEDs of which can be found in Appendix~\ref{app:SEDs}:
\begin{enumerate}
    \item When both IR and X-ray coverage are robust, the inclusion of the radio band has minimal effect. The $\chi^2$  remains largely unchanged, as do key parameters such as stellar mass, SFR, and AGN luminosity. In this case, adding the radio band provides little benefit at the expense of a large increase in the computational cost. The main advantage in this case is the ability to separate the AGN and SF-related emission in the radio band, although this could also be achieved by exploiting the FIR-radio correlation for the SF component \citep[e.g.,][]{delvecchio21} or the X-ray to radio correlation for the AGN \citep[e.g.,][]{damato22}.
    \item In cases where there is no detection in the X-ray and IR bands, or when IR coverage is sparse, the radio band becomes crucial for accurately constraining the AGN component. Without sufficient IR and X-ray data, the AGN emission is poorly constrained, and CIGALE tends to miss the AGN component, especially for obscured AGN where the flux at optical and near-IR wavelengths is dominated by the host galaxy emission. However, incorporating the radio band can reveal the presence of the AGN, with the radio excess (relative to pure star-forming emission) helping to constrain the AGN fraction and luminosity.
    \item In some instances where X-ray detection is lacking, the radio band also aids in constraining the SFR. This is especially true when the radio emission can be mostly ascribed to SF-related processes, without any significant radio-excess. In particular, we observed cases where, without the radio data, the code attempted to fit the entire IR spectrum with an overly luminous AGN component, resulting in sub-optimal fits, underestimation of the SFR, and, consequently, overestimation of the AGN bolometric power.
\end{enumerate}

\subsection{Main Sequence}\label{sec:main_sequence}
From the SED fitting, we are able to reliably obtain the main properties of the galaxies hosting our AGN. For the \nev\ sources, we found a median stellar mass of $\log{(M_{*}/\rm{M_{\odot}})} = 10.8_{-0.4}^{+0.3}$ and a star-formation rate of $SFR = 9_{-7}^{+16}\,\rm{M_{\odot}/yr}$. The \civ-selected AGN have similar stellar masses (but with larger dispersion) of $\log{(M_{*}/\rm{M_{\odot}})} = 10.9_{-1.3}^{+0.3}$ and higher star-formation rate $SFR = 21_{-13}^{+40}\,\rm{M_{\odot}/yr}$.\par
\begin{figure}
	\includegraphics[width=\columnwidth]{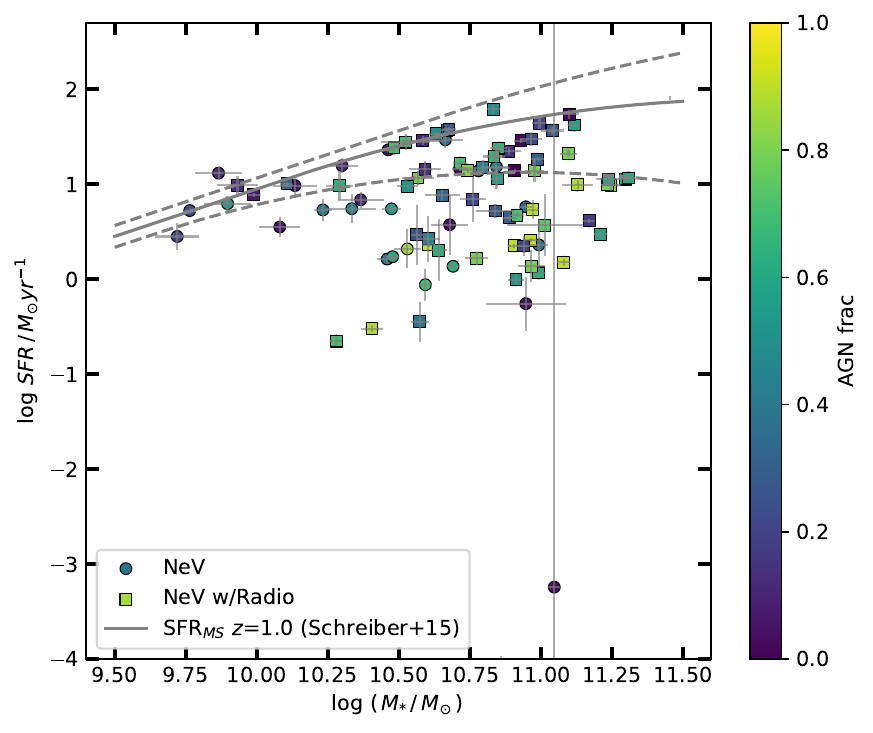}
    \caption{Comparison of the positions of the \nev\  sample in the $SFR-M_*$ plane. The squares represent the \nev\ sources detected in the radio band (either MeerKAT or VLA), while the circles represent those without radio detection. The colour code denotes the $1-1000\,\mu$m AGN fraction. The grey solid line is the \citet{schreiber15} MS at the mean redshift of the sources, the grey dashed lines its $1\,\sigma$ dispersion. The source with $\log{(M_*\,/\,\rm{M_{\odot}})}=11.04$ and $\log{(SFR_*\,/\,\rm{M_{\odot}\,yr^{-1}})}<-3$ is zCOSMOS ID 813366, for which we were not able to properly constrain the SFR (and the AGN properties) due to its complete lack of detections in the X-ray, radio, and IR bands. The source is therefore excluded from the rest of this work. } 
    \label{fig:ms_nev}
\end{figure}
\begin{figure}
	\includegraphics[width=\columnwidth]{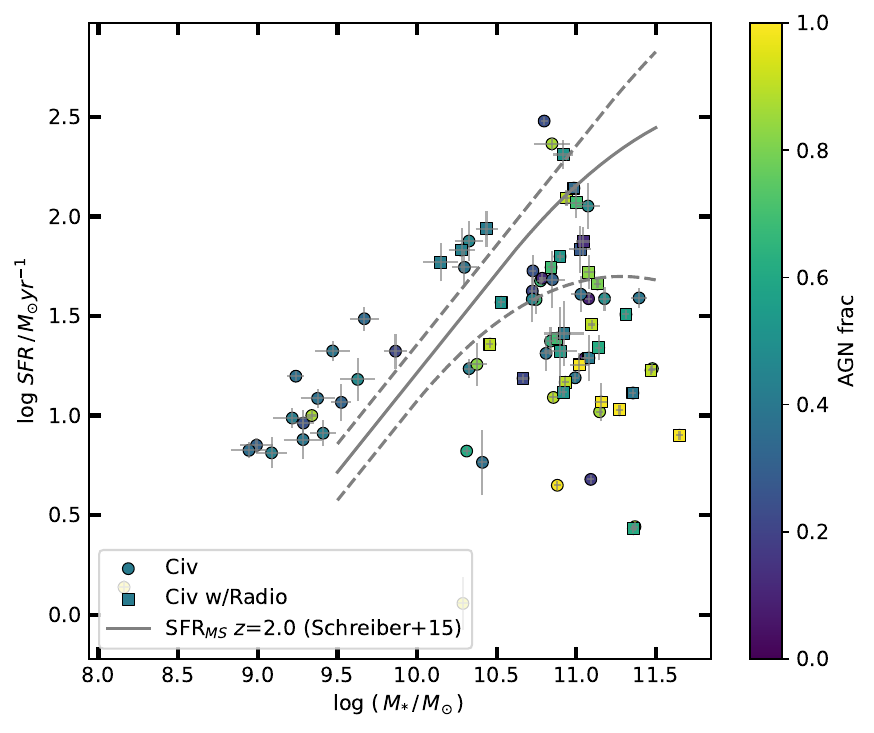}
    \caption{Comparison of the positions of the \civ\  sample in the$SFR-M_*$ plane. The squares represent the \civ\ sources detected in the radio band (either MeerKAT or VLA), while the circles represent those without radio detection. The colour code denotes the $1-1000\,\mu$m AGN fraction. The grey solid line is the \citet{schreiber15} MS at the mean redshift of the sources, the grey dashed lines its $1\,\sigma$ dispersion.} 
    \label{fig:ms_civ}
\end{figure}

We investigate whether our sources lie within the $SFR-M_{*}$ main sequence \citep{noeske07}, as shown in  Fig.~\ref{fig:ms_nev} for the \nev\ AGN and Fig.~\ref{fig:ms_civ} for the \civ\ sample. We use the \citet{schreiber15} main sequence (MS), in which the $SFR$ is a function of both the stellar mass and the redshift:
\begin{equation} 
\log{(SFR_{\text{MS}}/\text{M}_{\odot}\,\text{yr}^{-1}])}=\,m-m_0+a_0r-a_1[\max (0,m-m_1-a_2r)]^2
\label{eq:MS}
,\end{equation}
where $r\equiv \log(1+z)$, $m \equiv \log (M_{*}/10^9 \text{M}_{\odot})$, $m_0=0.5 \pm 0.07$, $a_0=1.5 \pm 0.15$, $a_1=0.3 \pm 0.08$, $m_1=0.36\pm 0.3,$ and $a_2=2.5 \pm 0.6$. \par


Among our sample, the source with  $\log{(M_*\,/\,\rm{M_{\odot}})}=11.04$ and $\log{(SFR_*\,/\,\rm{M_{\odot}\,yr^{-1}})}<-3$ corresponds to zCOSMOS ID 813366. For this object, we were unable to obtain reliable estimates of the SFR and the AGN properties, as it lacks any detections at X-ray, radio, or IR bands. Consequently, we excluded it from the analysis presented in the remainder of this work.\par
By comparing the position of each source with the MS at the source redshift, we find that most of the \nev\ sources are below the MS, while the \civ\ are more uniformly distributed, with $\sim15\%$ above the $1\sigma$ upper boundary. We also find that, for the \civ\, most of the radio detections lie at $\log{(M_*\,/\,\text{M}_{\odot})} \geq 10.5$. Finally, we note that the sources dominated by the AGN (i.e., those with the highest AGN fraction) tend to be segregated at high-mass and low $SFR$. \par
\begin{figure}
	\includegraphics[width=\columnwidth]{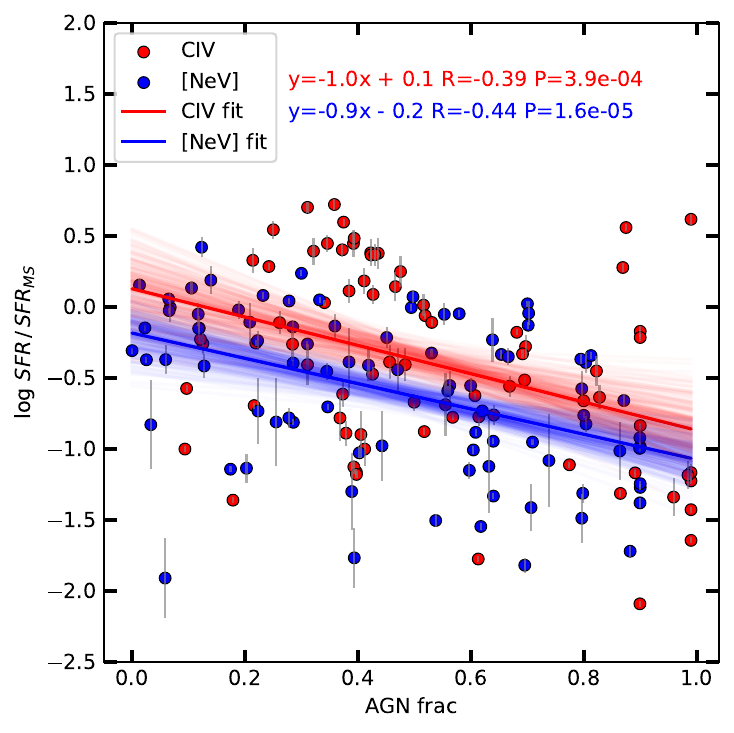}
    \caption{Offset with respect to the MS as a function of the AGN fraction for the \nev\ (\textit{blue}) and \civ\ (\textit{red}) samples. The \textit{solid} lines are the best-fit from linear regression analysis, the \textit{light} lines are obtained by bootstrapping the samples 300 times. R and P refer to the Pearson correlation coefficient and P-value, respectively.} 
    \label{fig:nev_civ_deltams_agnfrac}
\end{figure}

In Fig.~\ref{fig:nev_civ_deltams_agnfrac}, we plot the offset with respect to the main sequence $\Delta MS=\log{(SFR / SFR_{MS})}$ as a function of the AGN fraction for the \nev\ and \civ\ samples. $SFR_{MS}$ is computed at the redshift of each source. For both the \nev\ and \civ\ samples, we find the same trend seen in Figg.~\ref{fig:ms_nev} and \ref{fig:ms_civ}, with the sources with the highest AGN fractions being the farthest from the MS. The AGN fraction shows a stronger correlation with the $\Delta MS$ than with just the stellar mass (AGN fraction–$M_*$ Pearson $R \sim 0.2$), suggesting that the $\Delta MS$ trend is not merely driven by more luminous AGN residing in more massive galaxies, but may instead indicate that AGN activity has already quenched the SF in the sources farthest from the MS.

\subsection{$M_{\rm{BH}}$ and Eddington Ratio}
The Eddington ratio ($\eddrat=\text{L}_{\text{bol}}/\text{L}_{\text{Edd}}$) - the ratio between the AGN luminosity and the theoretical maximum luminosity that can be emitted by the AGN (determined by the balance between radiation pressure and the BH gravitational attraction) - is a useful indicator of the small-scale accretion state of an AGN. We used the \citet{suh20} $M_{\text{BH}} - \text{M}_{*}$ relation to estimate the SMBH masses using the stellar masses obtained from the SED-fitting. This relation was obtained from a sample of 100 $0.3<z<2.6$ X-ray selected AGN in the COSMOS field with host galaxy masses from SED-fitting decomposition and BH masses computed considering single epoch H$\alpha$, H$\beta$, and \mgii\ broad-line widths and line/continuum luminosity as a proxy for the size and velocity of the broad-line region (BLR). The \nev\ sample shows a uniform distribution of \mbh centred at  $\log{(M_{\rm{BH}}/\rm{M_{\odot}})}=7.3_{-0.6}^{+0.4}$, while the \civ\ AGN have a broader distribution centred at $\log{(M_{\rm{BH}}/\rm{M_{\odot}})} = 7.4_{-1.9}^{+0.4}$, clearly resembling their stellar mass distributions. We must caution against overinterpreting these $M_{\rm{BH}}$ values, as our current knowledge of $M_{\rm{BH}}-M_{*}$ relation is mostly local and usually derived from unobscured AGNs (due to the fact that the \mbh estimates are obtained from the profile of broad emission lines). Moreover, the advent of JWST has recently highlighted the presence of several high-$z$ galaxies with overmassive BH with respect to local relations \citep[e.g.,][]{ubler23,harikane23,pacucci24} and opened the possibility of a redshift evolution of the M$_{\text{BH}} - \text{M}_{*}$ relations. Finally, $\sim20\%$ of the \civ\ sources has $\log{(M_*/\,\rm{M_{\odot}})}<10$, which is outside the $10<\log{(M_*/\,\rm{M_{\odot}})}<12.5$ range in which this scaling relation has been calibrated. \par
\begin{figure}
	\includegraphics[width=\columnwidth]{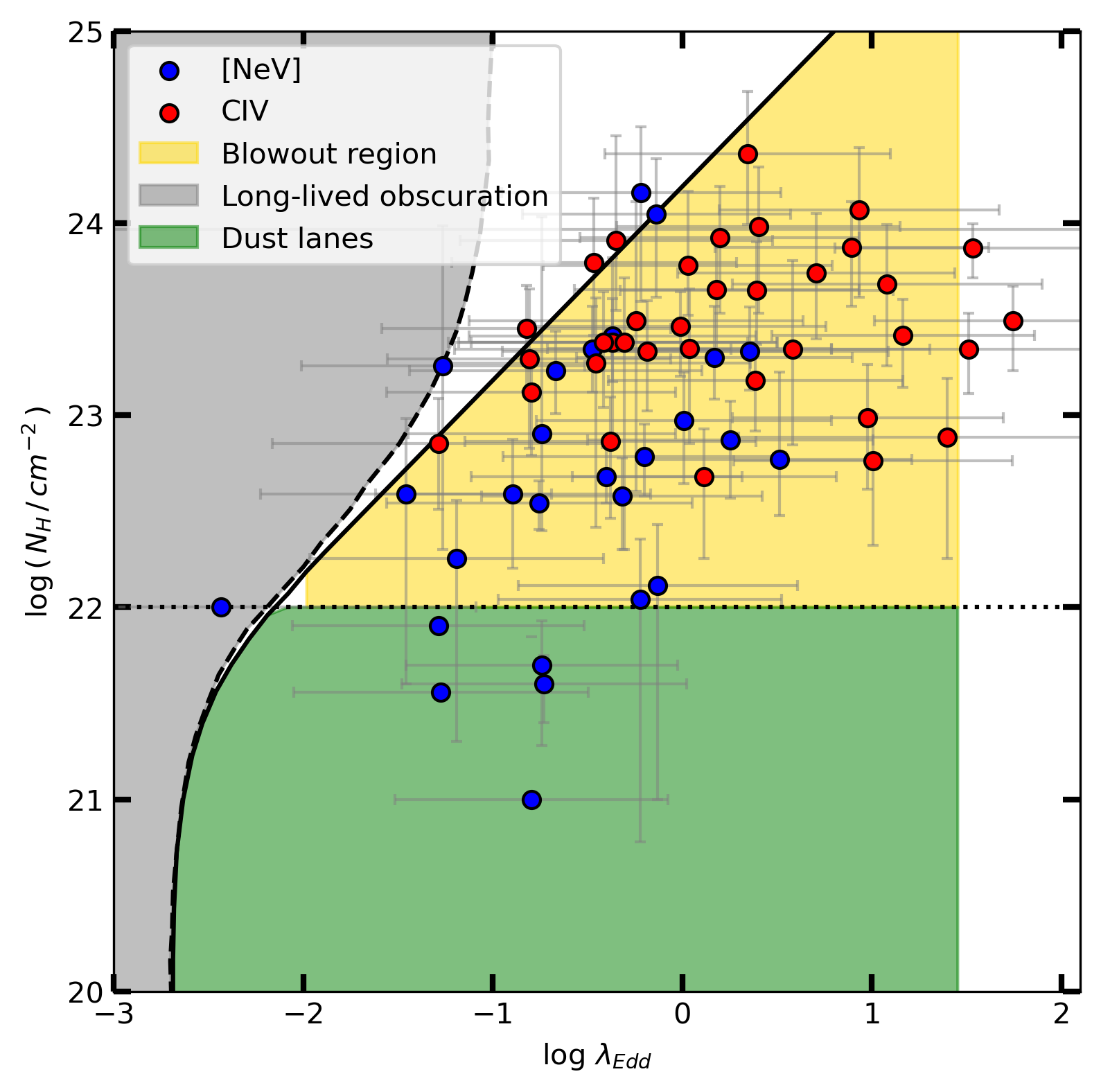}
    \caption{Amount of obscuration ($N_H$) as function of the Eddington ratio (\eddrat) for the \nev\ (\textit{blue}) and \civ\ (\textit{red}) AGN in our sample. The continuous (dashed) line is the effective Eddington limit $\lambda^{\rm{eff}}_{\rm{Edd}}$ for different values of column density based on the radiation-trapping (single-scattering) limit  \citep[e.g.][]{ricci18,ishibashi18}. The \textit{green} area is the region where the host galaxy could contribute to the AGN observed obscuration, while the \textit{yellow} one represents the blowout region, i.e., the locus where the AGN radiation pressure should push away the obscuring material. Finally, the \textit{grey} region identifies the locus where long-lived obscuration is possible.} 
    \label{fig:eddratio_nh}
\end{figure}

Due to the \civ\ sources having higher bolometric luminosities and similar or lower stellar masses than the \nev\ sample, it is not surprising to find that the \civ\ AGN have higher Eddington ratios.
In Fig.~\ref{fig:eddratio_nh}, we show the position of our sources in the $N_H$ versus Eddington ratio plot. Three different regimes are highlighted: the blowout region associated with short-lived obscuration, the long-lived obscuration region, and the region where obscuration is primarily due to host-galaxy dust lanes. The blowout region should be populated by AGN in a transient phase of their evolution, wherein the obscuring material will be blown away due to high radiation pressure emanating from the highly accreting AGN. On the contrary, obscuration can be long-lasting in AGN with a low accretion rate. A large fraction of the \nev\ AGN and most of the \civ\ lie on the blowout region, indicating that our sources are likely in a temporary phase of obscured accretion, which will soon end due to the AGN expelling most of its obscuring material.  Alternatively, as suggested by \citet{ballo14}, if the dust-to-gas ratio is lower than the typical of the ISM assumed by \citet{fabian09}, the radiation pressure would have a harder time pushing away the material, and the blowout region could move to higher $\eddrat$. In any case, we must be cautious in the interpretation of these results, as the Eddington ratio is heavily dependent on the BH mass, which, as already discussed, suffers from large uncertainties and possible biases. 

\subsection{Radio Properties}

\subsubsection{AGN and SF Radio contributions}
\begin{figure}
	\includegraphics[width=\columnwidth]{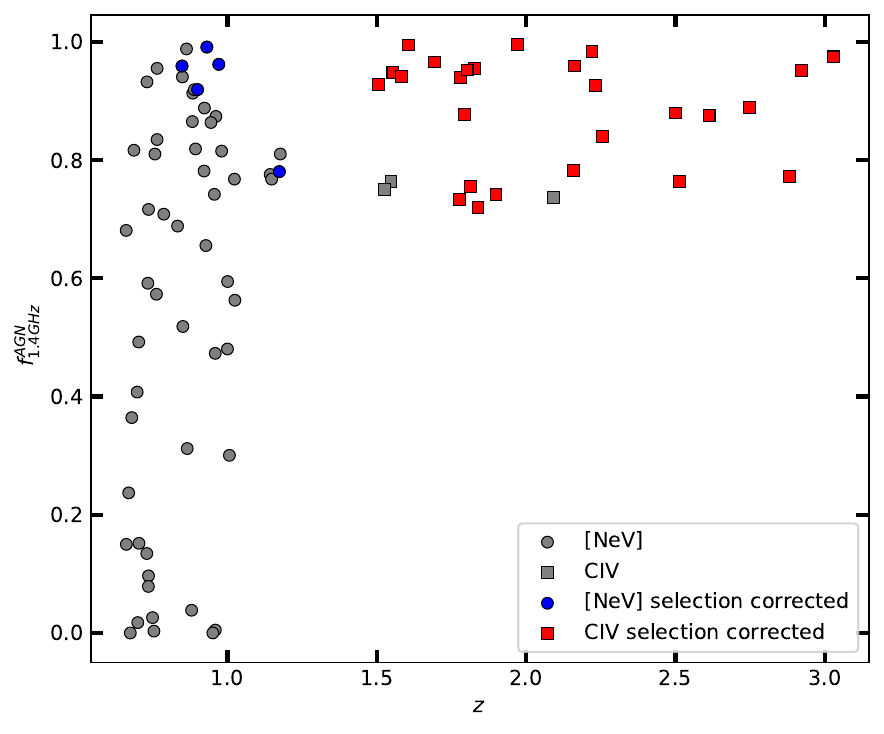}
    \caption{Fraction of the 1.4GHz radio luminosity that can be ascribed to AGN-related processes ($f_{1.4\text{GHz}}^{\text{AGN}}$) versus redshift for the \nev\ (\textit{circle}) and \civ\ (\textit{square}) sources. \textit{Blue} (\textit{Red}) points are \nev\ (\civ) ``selection corrected'' sources, i.e., those with  AGN luminosity $L_{1.4\rm{GHz}}>10^{40}\rm{erg/s}$ that can be detected in the entire redshift range. The lower number of \nev\ AGN with $L_{1.4\rm{GHz}}>10^{40}\rm{erg/s}$ is due to the smaller comoving volume.} 
    \label{fig:radio_agn_frac}
\end{figure}

The SED-fitting allows us to properly separate the AGN and SF contributions to the radio band flux. Overall, we find that our radio emission is dominated by the AGN component, with a median AGN $1.4\,$GHz fraction of $f_{1.4\text{GHz}}^{\text{AGN}}\equiv L_{1.4\text{GHz}}^{\text{AGN}} / (L_{1.4\text{GHz}}^{\text{AGN}} + L_{1.4\text{GHz}}^{\text{SF}}) = 0.8_{-0.5}^{+0.1}$. In particular, considering both samples and only the sources with radio detection, we have $13\%$ of the sample with a radio SF-dominated (i.e. $f_{1.4\text{GHz}}^{\text{AGN}}\leq 0.3$), $12\%$ with an almost equal contribution of SF and AGN, and $75\%$ with a radio AGN-dominated ($f_{1.4\text{GHz}}^{\text{AGN}}\geq 0.7$). Interestingly, as we can see from Fig.~\ref{fig:radio_agn_frac}, while the \nev\ sample has an almost uniform distribution in $f_{1.4\text{GHz}}^{\text{AGN}}$, the radio flux has a clear AGN origin for most of the \civ\ sample. This difference is due to selection effects. If we focus only on sources with $L_{\rm{1.4\,GHz}}\geq10^{40}\,\rm{erg/s}$ (i.e., the luminosity corresponding to the $5\sigma$ survey detection limit at $z=3$), we realise that most of the \nev\ AGN are below this threshold and cannot be detected at \civ\ redshifts. All the remaining five \nev\ AGN (i.e. those luminous enough to be detectable at $z=3$) have $f_{1.4\text{GHz}}^{\text{AGN}} \geq 0.8$, not dissimilar from \civ\ AGN. The low number of selection-corrected (i.e., $L_{\rm{1.4\,GHz}}\geq10^{40}\,\rm{erg/s}$) \nev\ AGN with respect to the \civ\ ones is linked to the $4.4$ times larger comoving volume for the \civ\ sample. In Fig.~\ref{fig:radio_agn_frac}, the \nev\ and \civ\ selection-corrected AGN are coloured, while the grey ones are those below the $L_{\rm{1.4\,GHz}}=10^{40}\,\rm{erg/s}$ threshold. 

\subsubsection{Radio$-$X-ray correlation}
\begin{figure}
	\includegraphics[width=\columnwidth]{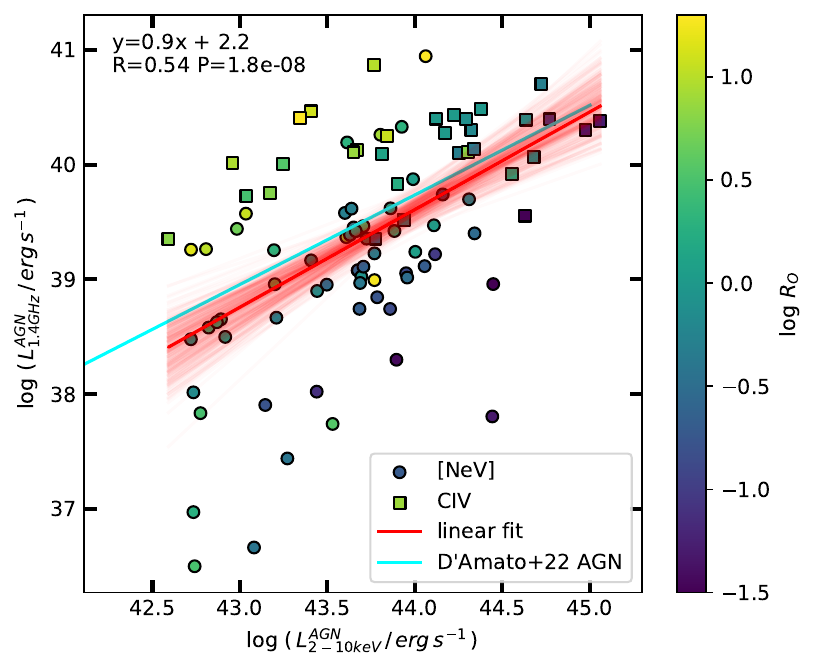}
    \caption{AGN $1.4\,$GHz luminosity versus rest frame $2-10\,$keV luminosity, for the \nev\ (\textit{circle}) and \civ\ (\textit{square}) sources with both X-ray and radio detection. The colour code indicates the optical radio loudness $R_O$. The \textit{dark red} line is the best-fit from linear regression analysis, and the \textit{light red} lines are obtained by bootstrapping the sample 300 times. The \textit{cyan} line is the same correlation found by \citet{damato22} for AGN in the J1030 field. R and P refer to the Pearson correlation coefficient and P-value, respectively.} 
    \label{fig:xray-radio_relation}
\end{figure}

In Fig.~\ref{fig:xray-radio_relation}, we show the position of our sources in the Radio$-$X-ray luminosity plane for the sources with both X-ray and VLA or MeerKAT detections. $L_{1.4\rm{GHz}}^{\rm AGN}$ refers to the AGN radio emission (i.e. without the SF contribution) as computed by CIGALE. As we can see, we find a very significant correlation (Null Hypothesis probability  $P_{\text{nh}}=11.8\times10^{-8}$) between the two quantities. This correlation is also extremely similar to the one from \citet{damato22}, which derived it from a sample of 63 AGN at $0<z<3$ (and two AGN at $z>3$) in the J1030 field. This correlation arise from the X-rays and radio emission having a common origin (i.e., the SMBH activity). As this correlation is likely affected by the underlying luminosity-distance correlation, we perform a Kendall $\tau$ correlation test of the three parameters: $L_{2-10keV}$, $L_{1.4\rm{GHz}}$, and $z$. We find a significant partial correlation between the radio and X-ray luminosity (once taking into account the redshift dependence) with $P_{\text{nh}}=3\times10^{-4}$, indicating that the $L_{2-10keV}-L_{1.4\rm{GHz}}$ is a real correlation and not only driven by the redshift evolution.  

\subsubsection{Radio Loudness}
We examined the radio loudness of our sample by adopting two radio loudness indicators: the optical radio loudness (defined in CIGALE as $R_O=L_{\nu,5\text{GHz}}^{\text{AGN}}/L_{\nu,2500\AA}^{\text{AGN,intr}}$ measured at an inclination $i=30$\textdegree), and the X-ray radio loudness $R_X=\log{(L_{1.4\text{GHz}}^{\text{AGN}}/L_{2-10\text{keV}}^{\text{AGN,intr}})}$. As both the AGN and SF components contribute to the radio luminosity, for $L_{\nu,5\text{GHz}}$ and $L_{1.4\text{GHz}}$, we choose to use the radio luminosities of the AGN component, obtained from the SED fitting. Finally, we assume a threshold for defining a source as a radio-loud AGN of $R_O>10$ \citep[e.g.][]{stoke92} and $R_X>-3.5$ \citep[e.g.][]{terashima03,lambrides20}.\par
Of the 100 sources with secure detection in the radio band, 16 (13) are classified as radio loud according to their $R_O$ ($R_X)$, and 12 are in common to both criteria. Notably, most of the radio loud sources are \civ\ AGN. In fact, independently of the chosen radio loudness criteria, the \civ\ sample shows a higher fraction of RL AGN ($\sim 25\%$) than the \nev\ sample ($\sim 6\%$), which is not surprising as the \civ\ sample is composed of more luminous sources and at higher redshift.  \par
\begin{figure}
	\includegraphics[width=0.85\columnwidth]{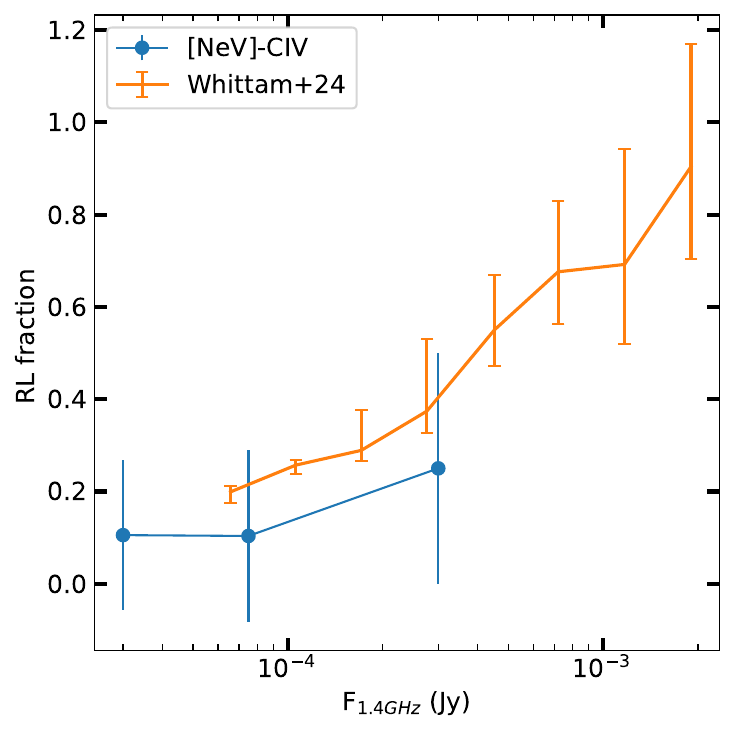}
    \caption{Comparison of the radio loudness fraction as a function of the source radio flux between this work (\textit{blue}) and the central $0.86\,\text{deg}^2$ of the COSMOS field by \citet[][\textit{orange}]{whittam24}. In both cases, we computed the flux densities at $1.4\,$GHz by rescaling the measured ones assuming a spectral index of $\alpha=0.7$.} 
    \label{fig:RL_fraction}
\end{figure}

In Fig.~\ref{fig:RL_fraction}, we compare the radio loudness fraction of our sources with the one computed by W24 in the central $0.86\,\text{deg}^2$ of the COSMOS field. For a fair comparison, we computed the RL fraction considering only the sources with a MIGHTEE detection (i.e., excluding the few sources with only VLA detection). Besides not having any radio bright source in our sample, we find a comparable radio loudness fraction to the one from the MIGHTEE survey. \par

We also investigated whether the properties of our RL-AGN differ from those of Radio Quiet (RQ) AGN. To this end, we compared the stellar mass, SFR, AGN fraction, and AGN bolometric luminosity of RL and RQ AGN in our sample. Defining the sources that satisfied either the optical radio loudness criteria or the X-ray ones as RL AGN, we found no significant difference between the two samples, with KS test P-values of $\sim0.5$ for all the properties. Interestingly, when we compared the same properties between the radio-detected and the radio-undetected sources, they are significantly different. In particular, the KS-test between the stellar mass distribution has a P-value of $\sim7\times10^{-9}$, with the radio-undetected sample having, on average, lower stellar masses than the radio-detected. In addition, both the AGN fraction and AGN bolometric luminosities of the two samples differ at $>99.5\%$ confidence. The latter could be explained by the fact that the AGN luminosity and fraction correlate with the host galaxy stellar mass.

\subsubsection{Radio Spectral Index}\label{sec:spec_index}
For the sources with secure detection both in MIGHTEE and VLA, we are able to derive their radio spectral index, simply defined as:
\begin{equation}
    \alpha=\frac{\log{(F_{\nu_1}/F_{\nu_2})}}{\log{(\nu_1\, /\, \nu_2)}}
\end{equation}
where $F_{\nu_1}$ is the MIGHTEE flux measured at the source effective frequency, and $F_{\nu_2}$ is the VLA $3\,$GHz flux from \citet{smolcic17a}.\par
From the 68 sources with detection in both bands, we obtain a median spectral index of $\alpha=-0.9 \pm 0.2$. 43 of these sources are \nev\ AGN with a median $\alpha=-0.9 \pm 0.2$, and the remaining 25 are \civ\ AGN, with an average spectral index of $\alpha=-1.0 \pm 0.2$. 5 sources (3 \nev\ and 2 \civ\ AGN) have $\alpha>-0.5$, while 7 (3 \nev\ and 4 \civ) have $\alpha<-1.2$. The presence of these Ultra-Steep Sources (USS) is consistent with previous works \citep{singh14}.  Two key factors must be considered when interpreting our measurements of the spectral index. First, due to the relatively long baselines of the VLA and its high spatial resolution (0.78''), we may lose some of the extended emission. Given that most of our sources are unresolved in MIGHTEE, this results in lower measured flux densities at 3 GHz compared to 1.3 GHz, leading to artificially steeper spectral indices. This is a possible factor contributing to our median spectral index being steeper but still consistent (considering the associated uncertainties) than the average $\alpha=-0.7$ of most AGN \citep{condon02}. Lastly, the five-year gap between the VLA and MIGHTEE observations must be considered, as intrinsic flux variability over this timescale could bias the inferred spectral index, making it appear steeper or flatter than its true value.\par
\begin{figure}
	\includegraphics[width=\columnwidth]{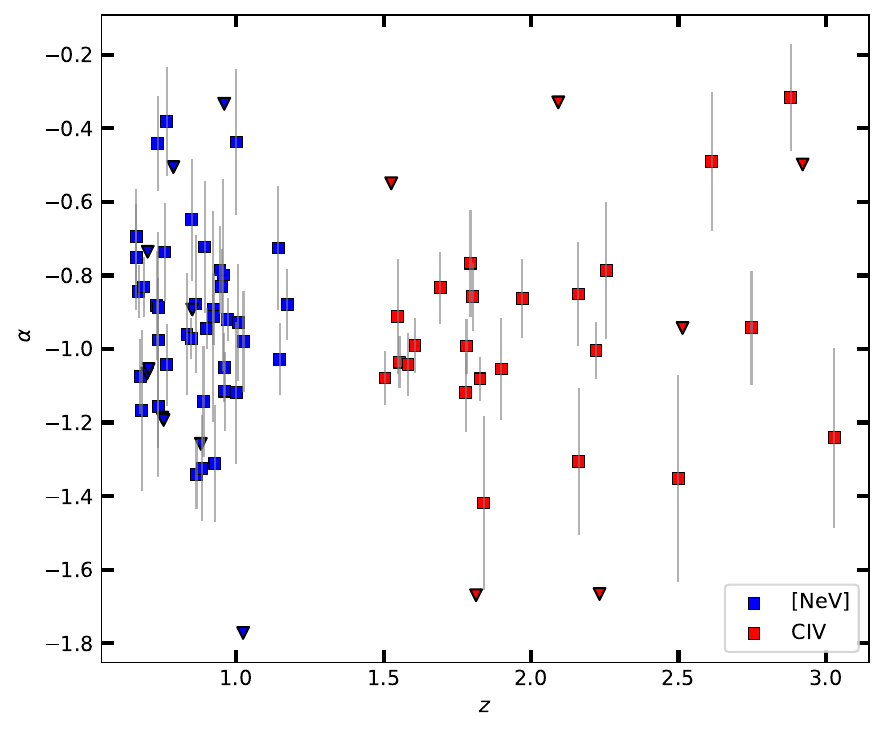}
    \caption{Radio spectral index as function of redshift for the \nev\ (\textit{Blue}) and for the \civ\ (\textit{Red}) AGN. The radio spectral index is measured using the $3\,$GHz flux from \citet{smolcic17a} and MIGHTEE flux at the effective frequency of each source (median $\nu_{\text{eff}}=1.24\,$GHz). Triangles denote upper limits, i.e. sources with MIGHTEE detection but undetected by VLA.} 
    \label{fig:spec_index}
\end{figure}

As shown in Fig.~\ref{fig:spec_index}, there is no significant difference in the spectral indices between the two samples, nor evidence of redshift evolution. For the 14 sources detected in MIGHTEE but not in the VLA, we used the VLA $5\sigma$ upper limits to compute the upper limits on their spectral indices. No clear trend was observed in these upper limits, and these sources are uniformly distributed in the same parameter space as those detected by the VLA.

\subsubsection{Properties of Radio undetected sources}
We conducted a comparative analysis of the primary characteristics of sources with radio detections (from either MIGHTEE or VLA) and those lacking such detections. No significant differences were observed in their spatial distribution or in their radio properties, aside from the expected lower radio flux in the undetected sources. However, the radio-undetected sources exhibit both lower AGN fractions and AGN luminosities, with statistical significances exceeding 99.5\% and 99.7\%, respectively (P-values from the KS test of $4.6 \times 10^{-3}$ and $2.3 \times 10^{-3}$). Additionally, although both samples occupy similar stellar mass ranges, the $M_*$ distribution of the radio-undetected sample is significantly distinct from that of the radio-detected sample, showing a skew towards lower masses, with a KS-test P-value of less than $10^{-8}$.
\begin{figure}
	\includegraphics[width=\columnwidth]{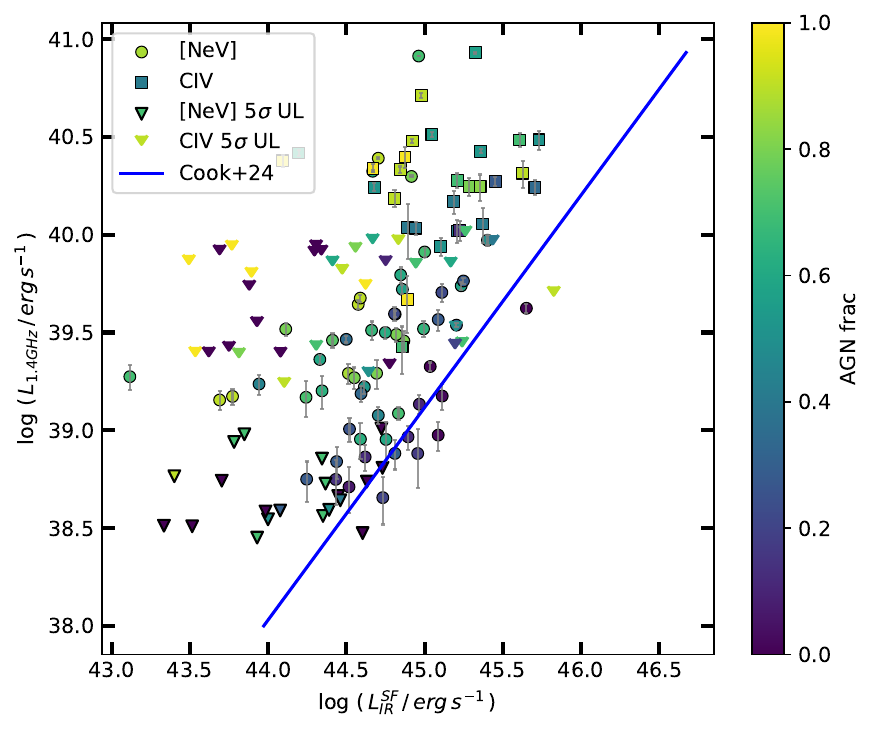}
    \caption{$1.4\,\rm{GHz}$ radio luminosity as function of the SF-related $1-1000\mu\rm{m}$ IR luminosity. The colour code indicates the AGN fraction in the $1-1000\mu\rm{m}$ wavelength range. $L_{1.4\rm{GHz}}$ is computed from the MIGHTEE $5.2''$ total flux, assuming spectral index $\alpha=-0.7$. The \textit{triangles} and \textit{arrows} indicate the $5\sigma$ upper limits on the radio luminosity, calculated at each source position from the MIGHTEE $5.2''$ sensitivity map, for the \nev\ and \civ\ sources, respectively. $L_{IR}^{SF}$ is an output of the SED-fitting procedure, and it takes into account only the IR emission due to SF processes, i.e. without any AGN contribution. The \textit{blue line} refers to the \citet{cook24} IR-radio correlation in case of pure SF processes.} 
    \label{fig:radio_excess}
\end{figure}
In Fig.~\ref{fig:radio_excess}, we show the distribution in radio luminosity of the \nev\ (\textit{circles}) and \civ\ (\textit{squares}) sources as a function of the IR luminosity due to SF. The $L_{1.4\rm{GHz}}$ was computed from the MIGHTEE $5.2''$ total flux, assuming a spectral index $\alpha=-0.7$. The \textit{triangles} and \textit{arrows}  denote the $5\sigma$ upper limit on the $L_{1.4\rm{GHz}}$, for the \nev\ and \civ\ sources respectively, obtained from the \textit{rms} map at each source position. The CIGALE-derived $L_{IR}^{SF}$ refers to the IR luminosity due to SF, without any AGN contribution. The \textit{blue line} represents the \citet{cook24} IR-radio correlation in case of pure SF processes. The colour code indicates the AGN fraction measured in the $1-1000\mu\rm{m}$ wavelength range. As we can see, most of our sources are above the \citet{cook24} correlation, as their radio luminosities have a strong AGN contribution. The sources nearest the correlation are those with a minimal AGN contribution (i.e., low AGN fractions). Almost all the \civ\ radio-undetected sources and more than half of the \nev\ ones could have a significant radio excess (with respect to the pure SF emission). This indicates that the non-detection of these sources in the radio band can not be attributed to a lack of AGN radio emission, but it is probably a consequence of the MIGHTEE survey sensitivity and the high-$z$ of our sources. For the (mostly \nev) sources located near the \citet{cook24} correlation, the radio emission appears to be primarily driven by SF, with the AGN contribution likely playing only a minor role.\par 
We found no significant differences when using the spectral index measured in Section~\ref{sec:spec_index} instead of adopting the fixed value of $\alpha=-0.7$. The main consequence of using the measured indices is a reduction of the sample to only those sources with both MIGHTEE and VLA detections.

\subsubsection{Comparison with the general population of MIGHTEE radio sources}
We compared the properties of our sources with those from W24, which analysed the counterparts of all MIGHTEE radio sources in the central $0.86\,\text{deg}^2$ of the COSMOS field. Our radio-detected \nev\ and \civ\ AGN form a subset of this catalogue (see section~\ref{sec:data_radio}). For the comparison, we focused on sources with MIGHTEE radio detections and restricted the W24 sources to those with $z_{\rm{best}}$ within our redshift range ($0.6<z<1.2$ and $1.5<z<3.1$). The redshift distribution comparison shows that, while our AGN represents a small fraction of MIGHTEE sources at lower redshifts, our sample constitutes the majority of W24 sources with available spectroscopic redshifts at $z>2$. Both samples cover similar stellar mass and SFR parameter spaces, with no significant differences in their distributions. The radio flux distribution of our sample peaks at lower values compared to W24, and our sample lacks sources with $F_{\nu}> 1.5\,\rm{mJy}$. This may reflect the effect of our selection, which favours highly accreting sources and thus biases the sample toward radiative-mode rather than jet-mode accretion \citep{heckman14}. In any case, our results are consistent with the expected number of sources above $1.5,\rm{mJy}$ (0.6), given the source density derived from the MIGHTEE sample.\par
To perform a morphological comparison between our galaxies and those in W24, we cross-matched both samples with the \citet{tasca09} catalogue, using a matching radius of $0.5''$ and the ``Tasca and Cassata'' morphological classification, which is based on the evaluation of the concentration index, asymmetry parameter, and Gini coefficient on HST-ACS images with a resolution of $0.03''/\textrm{pix}$. We find that approximately $50\%$ of our sample consists of spiral galaxies, $40\%$ are early-types, and $10\%$ are irregulars. Compared to W24, our sample contains twice the proportion of early-type galaxies and half as many irregular galaxies. 

\subsection{Comparison with parent samples and hydrodynamical simulations}\label{sec:simulations}

Due to the high redshift of the \civ\ sources and the challenges in detecting and resolving them in the FIR bands, we could not build a redshift- and stellar mass-matched
parent sample of non-active galaxies (see Appendix~\ref{app:parent}). Thus, to explore the differences between the \nev\ and \civ\ AGN populations and assess how their properties relate to the general galaxy population at comparable redshifts, we made use of the \simba\ hydrodynamical simulations presented in sec~\ref{sec:simba}. 

We started by building an ``analogue sample'', i.e. composed of \simba\ sources at the same redshift of our \nev\ and \civ\ AGN and with similar (within $\pm0.1$ dex) $M_*$, SFR, and black hole accretion rate ($\dot M_{\rm{acc}}$). For each \nev\ and \civ\ source, we select three analogue galaxies. When fewer than three matches were found, we gradually relaxed the selection criteria by widening the SFR bins (up to 0.5 dex), followed by the $M_*$ and $\dot M_{\rm{acc}}$ bins (up to 0.3 dex). For 34 of our sources, no suitable analogues could be identified, which we primarily attribute to a high Eddington ratio ($\lambda_{\rm{Edd}}\gtrsim0.5$): such highly accreting galaxies are rare in the \simba\ simulation, making it difficult to simultaneously match $M_*$ and SFR. In addition, some of our highest-$\dot M_{\rm{acc}}$ sources exceed the $\eddrat<3$ limit imposed in \simba\ and are therefore absent altogether. Once we identified suitable analogues, we were able to trace their past and future evolutions, as each galaxy is linked to its most massive progenitor (descendant) in the previous (next) snapshot. \par
\begin{figure}
	\includegraphics[width=\columnwidth]{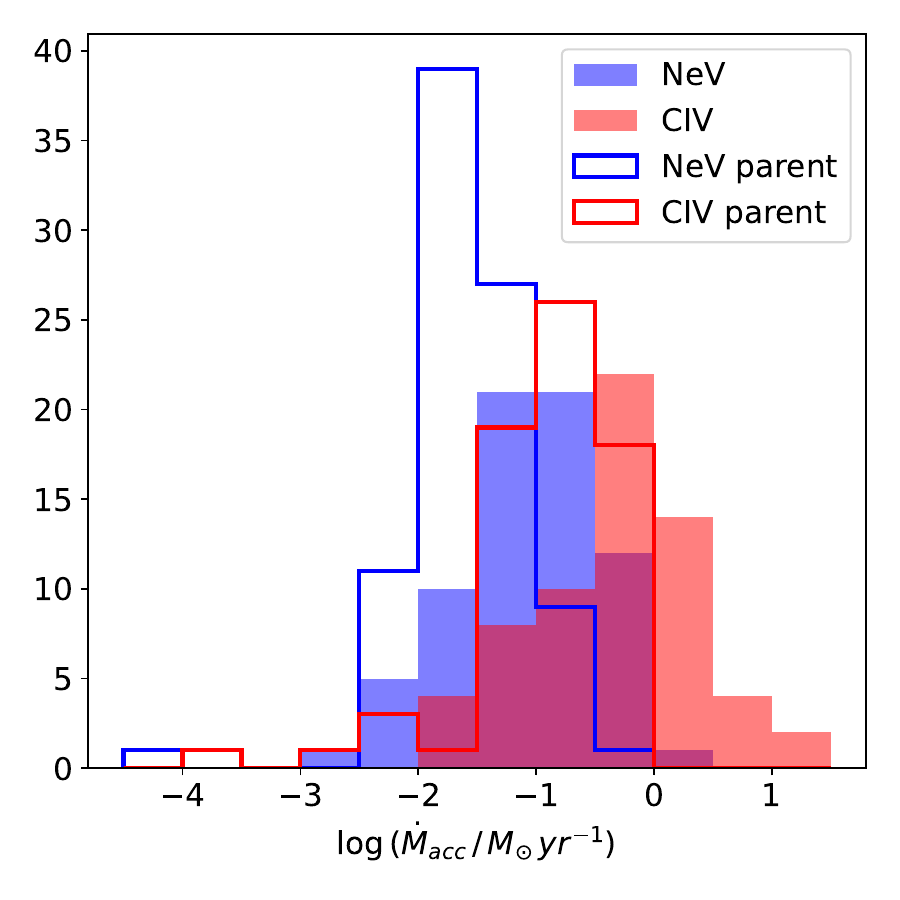}
    \caption{Distribution of the BH accretion rates of the \nev-\civ\ ``analogue sample'' and of their ``parent samples''. \textit{Blue} histograms are for the \nev\ AGN, \textit{red} ones refers to the \civ\ AGN. The ``analogue sample'' is composed of \simba\ galaxies matched in redshift, stellar mass, SFR, and $\dot M_{BH}$ to the \nev\ and \civ\ sources. The ``parent sample'' has been similarly built without matching $\dot M_{BH}$, representing the average galaxy population. It is easy to see that our selection picks out sources in a stronger-than-average phase of accretion.} 
    \label{fig:simba_macc}
\end{figure}

Additionally, we built a ``parent sample'' by selecting five \simba\ galaxies matched in $M_*$ and SFR, but not in accretion rate, for each \nev-\civ\ AGN. As shown in Fig.~\ref{fig:simba_macc}, the ``analogue sample'' systematically exhibits higher $\dot M_{\rm{acc}}$ than the ``parent sample'', indicating that our selection preferentially identifies galaxies in a strong phase of accretion. \par

\begin{figure*}
	\includegraphics[width=\textwidth]{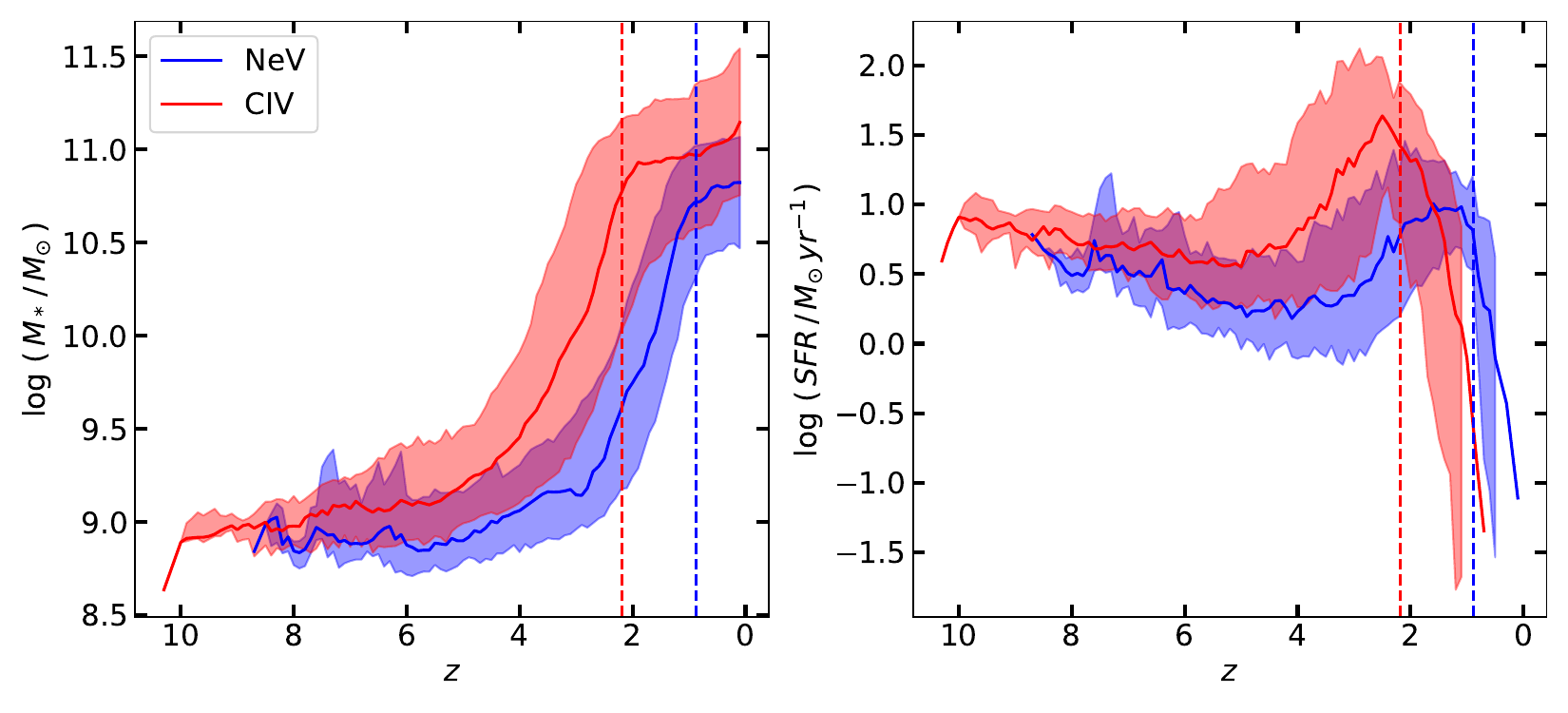}
    \caption{Average redshift evolution of the stellar mass (\textit{left panel}) and SFR (\textit{right panel}) of the \nev-\civ\ analogue galaxies in \simba\ simulations. \textit{Solid lines} represent the median values, while the \textit{coloured areas} encompass their $16^{th}-84^{th}$ percentiles. The \textit{dotted lines} indicate the median redshift at which our sources were observed for the \nev\ (\textit{blue}) and \civ\ (\textit{red}), respectively. } 
    \label{fig:simba_evolutions}
\end{figure*}

Figure~\ref{fig:simba_evolutions} presents the past and future evolution of the \nev\ and \civ\ analogues. The median redshifts at which the \nev-\civ\ sources have been observed are marked by dashed lines. As we can see, the evolutions of the \nev\ and \civ\ analogues are significantly distinct.  The \civ\ analogues began assembling their stellar mass earlier and exhibit a peak in their SFH at $z \sim 2.5$, whereas the \nev\ analogues grew later, with their SFH peaking at $z \sim 1.5$. Although their $M_*$ are comparable at the time of observation ($z_{\rm obs}$), the \civ\ analogues will evolve into significantly more massive systems by $z = 0$. Interestingly, both populations show SFH peaks slightly preceding their $z_{\rm obs}$, consistent with ongoing or recently completed star formation bursts. \par

When comparing the overall evolutions of the ``analogue samples'', we found no significant difference. This suggests that the \nev\ and \civ\ galaxies are not fundamentally different from the general population, but are rather observed during a transient phase of enhanced nuclear activity (see also Fig.~\ref{fig:simba_macc}), likely due to our selection targeting galaxies in this phase. This is in agreement with the position of the sources in the $N_H-\lambda_{\rm{Edd}}$ plot (Fig.~\ref{fig:eddratio_nh}): most of our sources are in the blowout region, and they should soon expel most of their obscuring material (and possibly transition to a phase of lower accretion). Previous studies have also suggested that high-ionisation lines, particularly the \nev\ one, may trace galaxies in a specific and transient phase of their evolution. For instance, \citet{vergani18} analysed 529 \nev\ emitters in the VIMOS Public Extragalactic Redshift Survey \citep[VIPERS;][]{guzzo14,garilli14,scodeggio18} and found that many are not identified through standard AGN selection methods (shallow X-ray data, mid-IR colours, or optical line diagnostics), suggesting that \nev\ may reveal a transient evolutionary phase, with a subsample of those sources showing indications of a sudden suppression of the star-formation with timescales of $200-300\,\text{Myr}$. Similarly, \citet{yuan16}, studying 2758 $z<1$ type 2 QSOs (including a subsample with strong \nev\ emission), proposed that the sources with \nev\ emission may be extremely obscured even at IRAC 1 and 2 wavelengths, which may explain why IR colour selection misses a significant fraction of them. \par
As discussed in Section~\ref{sec:main_sequence}, the \civ-selected AGN exhibit signs of bi-modality. In particular, sources located significantly below the MS tend to display higher AGN fractions, elevated \eddrat, and stronger radio emission. We explored whether we could find any difference in the evolution of these two putative sub-populations in \simba. We divided the \civ\ analogues into ``quiescent'' and ``star-forming'' subsamples based on their specific star formation rate ($sSFR$), adopting a threshold of $\log{(sSFR/\rm{yr}^{-1})}=-10$, which corresponds to an offset from the MS $>1$ dex. We found that the past and future evolution of the two sub-samples are significantly different: ``quiescent'' sources began building their stellar mass earlier ($z \sim 4.5$), and will evolve into systems with stellar masses $\sim 0.5-1$ dex higher at $z = 0$ compared to their ``star-forming'' counterparts, which started assemble their mass at $z \sim 3.5$. Their SFHs are also different: the ``quiescent'' sub-sample peaks at $z \sim 4$ with an average SFR of $\sim 100\, \rm M_\odot/yr$, while the ``star-forming'' SFH peaks at $z < 3$ with an average SFR of $\sim 30\,\rm M_\odot / yr$. Finally, the $SFR$ of the ``quiescent'' sources drops to zero soon after $z_{obs}$, while the ``star-forming'' galaxies continue to form stars (although at $SFR<10\,\rm{M_{\odot}/yr}$) up to $z\sim1$. 

\subsection{\nev\ and \civ\ lines select different populations}\label{sec:disc_diff_pop}
The \nev\ sample has been the focus of several studies \citep[e.g.][]{mignoli13,vignali14,barchiesi24}. In particular, B24 found that the \nev\ population is largely composed of obscured AGN (with $\sim20\%$ candidate CT-AGN), with about half exhibiting significant SMBH growth ($\lambda_{\rm{Edd}}>0.1$). While their hosts contain gas reservoirs comparable to those of inactive galaxies, they show systematically higher SFRs. Based on the \textit{in-situ} co-evolution model, these results were interpreted as evidence that \nev\ AGN are in a ``pre-quenching phase,’’ with only the older systems displaying signatures of AGN-driven suppression of the star formation. On the contrary, the \civ\ properties suggest that at least a significant fraction of the \civ\ AGN may be quenched by their AGN, which seems confirmed by the predicted evolution of their ``analogues'' in the \simba\ simulations (Sec.~\ref{sec:simulations}).\par
Regarding their properties, the \nev\ sources are broadly distributed around and below the MS, with their radio emission likely reflecting a mix of AGN activity and SF (Fig.~\ref{fig:radio_agn_frac}). In contrast, \civ\ AGN display a bimodal distribution in the $SFR-M_*$ plane (Fig.~\ref{fig:ms_civ}), and their radio emission is almost entirely AGN-dominated. Furthermore, the \simba\ analogues of the two samples show markedly different evolutionary paths, including the redshift at which the host galaxy begins to grow, the shape of the SFH, and their final stellar masses at $z=0$.\par
This seems to suggest that the selection via \nev\ and \civ\ lines identifies two different galaxy populations, both observed at or near the peak of their SFHs, but that will evolve into galaxies with markedly different properties. \par

\section{Conclusion}\label{sec:conclusion}
In this work, we constructed and analysed the first sample of narrow-line selected AGN with full coverage from X-ray to radio wavelengths. We exploited the high-ionisation emission lines \Nev\ and \Civ\ to identify obscured AGN within the redshift ranges $0.6<z<1.2$ and $1.5<z<3.1$, respectively. Our sample consisted of 184 sources located in the COSMOS field. By leveraging data from the \textit{Chandra} COSMOS Legacy survey, the COSMOS2020 catalogue, the XID+ deblended catalogue, and radio observations from VLA and MIGHTEE, we reconstructed the SEDs of our sources across the entire spectrum from the X-rays to the radio band. Using the CIGALE SED-fitting code, we disentangled the AGN and host-galaxy emissions, allowing us to characterise key properties such as stellar mass, SFR, AGN fraction, bolometric luminosity, and radio characteristics with unprecedented accuracy. We also investigate the possible past and future evolution of our sources by comparing with the results from \simba\ hydrodynamical simulations. We highlight the crucial role of radio data in SED fitting, with radio observations essential for accurately disentangling AGN and host-galaxy components, especially in cases where X-ray or IR data are sparse or absent. Our main findings are the following:
\begin{enumerate}
    \item Approximately $60\%$ (MeerKAT) and $40\%$ (VLA) of our sources have a radio counterpart, with $17\%$ classified as RL AGN. Their radio emission is a combination of AGN and SF-related processes for the \nev\ sample, while the radio emission of \civ\ AGN is mostly of AGN origin. 
    
    \item  Although RL sources exhibited higher radio luminosity, there are no significant differences between RL and non-RL sources regarding AGN or host-galaxy properties. However, AGN with radio detections, in general, showed higher stellar masses and higher AGN fractions compared to those without radio detections.

    \item The comparison with \nev\ and \civ\ analogues in \simba\ hydrodynamical simulations indicates that our selection is targeting a transient phase in galaxy evolution in which the sources are extremely active, rather than a particular class of objects with its own evolution. We put forward that our selection is indeed selecting galaxies in the elusive obscured accretion phase of their evolution.

    \item Our comparison with \simba\ simulations suggests that \civ\ sample is composed of two distinct sub-populations: the ``quiescent'' population (composed of galaxies with lower sSFR) has a higher AGN fraction, has formed earlier, with a stronger burst of SF, will form more massive galaxies at $z=0$, and its SFR is already quenched. The ``star-forming'' population shows lower AGN fractions, has assembled most of its stellar mass at later times, and will continue to form stars up to $z\sim1$.

    \item \nev\ and \civ\ line selections target two different AGN populations: the \civ\ sources have formed earlier, their SFH peaked at higher redshifts, and they will evolve into higher mass galaxies at $z=0$, while the \nev\ population formed later and in a less strong burst of SF. Furthermore, the \nev\ sources appear to be in a ``pre-quenching'' phase with an observed SFR not significantly different from the SFR at the peak of their SFH. Whereas the \civ\ sources, characterised by higher AGN luminosities, obscuration, and AGN-dominated radio emission, seem to have already been quenched by their AGN, especially those in the ``quiescent'' subpopulation.
    
\end{enumerate}

This work highlights the importance of extending the SED fitting beyond the traditional UV-to-FIR range. We demonstrated that incorporating the X-ray and radio bands is crucial for deriving reliable properties of AGNs and their host galaxies. Moreover, upcoming large-area spectroscopic surveys will provide spectra of millions of high-$z$ galaxies, making it possible to use these high-ionisation lines to select large samples of obscured AGN. For instance, the Wide-Area VISTA Extragalactic Survey \citep[WAVES,][]{driver16,driver19} and the Optical, Radio Continuum, and HI Deep Spectroscopic Survey \citep[ORCHIDSS,][]{duncan23} will target more than 180,000 galaxies, while the Wide Field Spectroscopic telescope (WFS), is expected to deliver $\textrm{R}=3000-4000$ Multi-Object Spectroscopy for 250 million sources, along with over 4 billion spectra via its $\textrm{R}=3500$ Integral Field Spectrograph \citep{bacon24}. Instruments such as MOONS and surveys like MOONRISE \citep{maiolino20} will allow us to further extend the use of the \nev\ and \civ\ lines to select obscured AGN up to $z\sim5$, making these works a crucial step toward fully uncovering and characterising the obscured AGN population.

\section*{Acknowledgements}
LB, LM, MV acknowledge financial support from the Inter-University Institute for Data Intensive Astronomy (IDIA), a partnership of the University of Cape Town, the University of Pretoria and the University of the Western Cape, and from the South African Department of Science and Innovation’s National Research Foundation under the ISARP RADIOMAP+ Joint Research Scheme (DSI-NRF Grant Number 150551) and the CPRR HIPPO projects (DSI-NRF Grant Number SRUG22031677, SRUG2204254729). VS acknowledge that the research work carried out at the Physical Research Laboratory is funded by the Department of Space, Government of India. The MeerKAT telescope is operated by the South African Radio Astronomy Observatory, which is a facility of the National Research Foundation, an agency of the Department of Science and Innovation. We acknowledge the use of the ilifu cloud computing facility – www.ilifu.ac.za, a partnership between the University of Cape Town, the University of the Western Cape, Stellenbosch University, Sol Plaatje University and the Cape Peninsula University of Technology. The Ilifu facility is supported by contributions from the Inter-University Institute for Data Intensive Astronomy (IDIA – a partnership between the University of Cape Town, the University of Pretoria and the University of the Western Cape, the Computational Biology division at UCT and the Data Intensive Research Initiative of South Africa (DIRISA). The authors acknowledge the Centre for High Performance Computing (CHPC), South Africa, for providing computational resources to this research project. CLH acknowledge support from the Oxford Hintze Centre for Astrophysical Surveys, which is funded through generous support from the Hintze Family Charitable Foundation.  IP acknowledges financial support from the Italian Ministry of Foreign Affairs and International Cooperation under the ``Progetti di Grande Rilevanza'' scheme (project RADIOMAP, grant number ZA23GR03) (DSI-NRF Grant Number 150551). RG and MM acknowledge support from the INAF 2022/2023 ``Ricerca Fondamentale'' grants. 

\section*{Data Availability}
The data underlying this article will be shared on request sent to the corresponding author.



\bibliographystyle{mnras}
\bibliography{nevciv_mightee} 




\appendix

\section{X-ray properties of the \civ\ sample} 
In table~\ref{tab:civ_xray_prop_all}, we report the X-ray properties of the \civ-selected AGN with X-ray detection.

\begin{table*}
\caption{X-ray properties of the X-ray detected \civ\ sources derived from the spectral analysis. The model consists of a power-law with fixed photon index $\Gamma=1.8$, a neutral absorber at the redshift of the source to represent the torus obscuration, and a local neutral absorber with $N_{H,\rm{gal}}\sim1.8\times10^{20}\,\rm{cm^{-2}}$ to take into account the absorption from our Galaxy. \textit{ID} is the zCOSMOS ID for our sources, $z$ is the redshift, \textit{Counts} represent the number of net (i.e. background-subtracted) counts. $L_{X,\rm{intr}}$ is the intrinsic (i.e., absorption corrected) $2-10\,\rm{keV}$ luminosity, and $ N_H $ is the amount of obscuration in units of $10^{22}\,\rm{cm^{-2}}$. \textit{Bkg} indicates a source for which we opted to model the background (see Sec.~\ref{sec:xray_analysis}) instead of subtracting it.}
\label{tab:civ_xray_prop_all}
\centering
\begin{tabular}{cccccc}
\hline 
\textit{ID} & $z$ & \textit{Counts} & $\log L_{X,\rm{intr}}$ & $ N_H $ & \textit{Bkg} \\  
\hline 
400615 & 2.2551 & 15 & $43.6 \pm 0.2$ &$19_{-16}^{+22}$ &  \\[2pt] 
410572 & 3.0261 & 43 & $44.8_{-0.2}^{+0.3}$ &$48_{-30}^{+50}$ &  \\[2pt] 
415416 & 2.4167 & 67 & $44.5 \pm 0.2$ &$28_{-15}^{+19}$ &  \checkmark  \\[2pt] 
413797 & 2.5899 & 166 & $44.5 \pm 0.1$ &$7.7_{-5.9}^{+7.9}$ &  \\[2pt] 
409359 & 1.6922 & 26 & $43.9 \pm 0.2$ &$24_{-16}^{+42}$ &  \checkmark  \\[2pt] 
406766 & 1.7776 & 49 & $44.2 \pm 0.2$ &$81_{-46}^{+203}$ &  \\[2pt] 
416133 & 2.0795 & 145 & $44.5_{-0.1}^{+0.0}$ &$5.7_{-2.7}^{+3.1}$ &  \\[2pt] 
412437 & 1.7655 & 38 & $43.1_{-0.8}^{+0.4}$ &$20_{-13}^{+26}$ &  \checkmark  \\[2pt] 
414241 & 1.8095 & 25 & $43.8 \pm 0.2$ &$24_{-22}^{+28}$ &  \\[2pt] 
435276 & 2.1653 & 12 & $43.1_{-0.6}^{+0.4}$ &$ <600 $ &  \checkmark  \\[2pt] 
411173 & 2.1517 & 26 & $43.9_{-0.3}^{+0.4}$ &$31_{-27}^{+98}$ &  \\[2pt] 
410027 & 1.8387 & 25 & $43.2 \pm 0.2$ &$ <0.1 $ &  \checkmark  \\[2pt] 
402019 & 2.3861 & 56 & $44.0 \pm 0.1$ &$ <4.6 $ &  \checkmark  \\[2pt] 
490806 & 2.1745 & 72 & $44.1 \pm 0.1$ &$ <6.9 $ &  \checkmark  \\[2pt] 
414412 & 2.1915 & 85 & $44.4 \pm 0.1$ &$22_{-9}^{+12}$ &  \\[2pt] 
411773 & 2.9750 & 31 & $44.5_{-0.2}^{+0.3}$ &$60_{-27}^{+87}$ &  \\[2pt] 
490296 & 1.6490 & 71 & $43.9 \pm 0.1$ &$4.8_{-3.0}^{+3.7}$ &  \\[2pt] 
400940 & 2.6125 & 67 & $44.9 \pm 0.1$ &$74_{-22}^{+25}$ &  \\[2pt] 
415518 & 2.9210 & 77 & $43.6_{-0.6}^{+0.3}$ &$ <19.2 $ &  \checkmark  \\[2pt] 
413569 & 2.7408 & 22 & $44.4_{-0.3}^{+0.1}$ &$85_{-48}^{+75}$ &  \\[2pt] 
409532 & 1.6388 & 33 & $43.9_{-0.2}^{+0.1}$ &$24_{-13}^{+20}$ &  \\[2pt] 
411643 & 1.5256 & 22 & $43.8 \pm 0.3$ &$22_{-15}^{+23}$ &  \checkmark  \\[2pt] 
410112 & 2.9412 & 64 & $43.3_{-1.0}^{+0.4}$ &$10_{-9}^{+11}$ &  \checkmark  \\[2pt] 
406142 & 1.5044 & 47 & $44.5 \pm 0.1$ &$55_{-30}^{+57}$ &  \checkmark  \\[2pt] 
406139 & 1.8132 & 163 & $44.5 \pm 0.1$ &$5.8_{-3.7}^{+4.3}$ &  \\[2pt] 
411563 & 1.7964 & 216 & $44.6_{-0.1}^{+0.0}$ &$7.3_{-4.4}^{+5.1}$ &  \\[2pt] 
403218 & 1.5472 & 37 & $43.9_{-0.4}^{+0.3}$ &$ <148 $ &  \\[2pt] 
401129 & 2.0922 & 92 & $44.5 \pm 0.1$ &$15_{-7}^{+8}$ &  \\[2pt] 
405121 & 2.2204 & 31 & $44.5_{-0.4}^{+0.3}$ &$117_{-76}^{+131}$ &  \checkmark  \\[2pt] 
413208 & 3.0285 & 191 & $45.0 \pm 0.1$ &$31_{-14}^{+16}$ &  \\[2pt] 
411494 & 2.5992 & 26 & $43.8_{-0.3}^{+0.2}$ &$ <57 $ &  \checkmark  \\[2pt] 
410299 & 1.9500 & 47 & $44.3 \pm 0.1$ &$29_{-13}^{+15}$ &  \\[2pt] 
404703 & 2.2337 & 40 & $43.8_{-0.1}^{+0.3}$ &$62_{-36}^{+73}$ &  \\[2pt] 
411270 & 2.2545 & 32 & $44.1_{-0.2}^{+2.4}$ &$30_{-19}^{+26}$ &  \\[2pt] 
405213 & 1.8991 & 54 & $45.0_{-0.4}^{+0.5}$ &$230_{-132}^{+256}$ &  \checkmark  \\[2pt] 
409858 & 1.6612 & 59 & $43.7 \pm 0.1$ &$ <3.6 $ &  \checkmark  \\[2pt] 
403380 & 2.1616 & 44 & $44.2 \pm 0.2$ &$22_{-15}^{+42}$ &  \\[2pt] 
411511 & 1.8343 & 72 & $43.3_{-1.0}^{+0.4}$ &$30_{-10}^{+13}$ &  \checkmark  \\[2pt] 
490304 & 1.5395 & 59 & $43.8 \pm 0.1$ &$9.7_{-5.6}^{+8.6}$ &  \\[2pt] 
409999 & 2.9465 & 43 & $44.5 \pm 0.2$ &$84_{-50}^{+71}$ &  \\[2pt] 
413776 & 1.5824 & 62 & $43.8 \pm 0.2$ &$26_{-12}^{+14}$ &  \\[2pt] 
409336 & 2.4419 & 45 & $44.4 \pm 0.1$ &$41_{-20}^{+25}$ &  \\[2pt] 
415387 & 2.1596 & 67 & $43.5_{-0.6}^{+0.4}$ &$44_{-21}^{+36}$ &  \checkmark  \\[2pt] 
402666 & 1.6003 & 60 & $43.7 \pm 0.1$ &$ <2.2 $ &  \checkmark  \\[2pt] 
411922 & 2.9619 & 36 & $44.0 \pm 0.2$ &$45_{-24}^{+34}$ &  \\[2pt] 
410011 & 2.2000 & 38 & $43.4_{-0.2}^{+0.3}$ &$21_{-15}^{+28}$ &  \\[2pt] 
406358 & 1.8574 & 77 & $43.2_{-0.7}^{+0.3}$ &$7.1_{-3.9}^{+5.1}$ &  \checkmark  \\[2pt] 
405772 & 2.2322 & 23 & $44.4 \pm 0.2$ &$96_{-62}^{+100}$ &  \checkmark  \\[2pt] 
402989 & 1.7605 & 48 & $43.8 \pm 0.2$ &$ <8.7 $ &  \checkmark  \\[2pt] 
406749 & 2.0472 & 30 & $44.3 \pm 0.2$ &$75_{-38}^{+54}$ &  \\[2pt] 
490342 & 3.0173 & 108 & $44.9 \pm 0.1$ &$13_{-7}^{+8}$ &  \checkmark  \\[2pt] 
401460 & 1.5536 & 108 & $44.1 \pm 0.1$ &$22_{-11}^{+18}$ &  \\[2pt] 

\hline
\end{tabular}
\end{table*}

\section{\civ\ parent sample of non-active galaxies}\label{app:parent}
Similarly to the approach adopted by B24 for the \nev\ sample, to further investigate the nature of the \civ\ sources, we aimed to compare those with the average population of non-active galaxies. To this end, we constructed and analysed a parent sample of ``normal'' galaxies, matched in redshift and stellar mass. However, within the COSMOS field, galaxies in the redshift range $1.6 < z < 3.1$ generally lack detections in the FIR bands. As a result, the SED-fitting procedure was unable to robustly constrain their properties, particularly the SFR. Conversely, galaxies with FIR detections were typically fitted with high AGN fractions, suggesting that at the current FIR survey depths, an AGN component is required to reproduce the observed emission. This indicates that the parent sample was also significantly contaminated by AGN activity. Therefore, we had to rely on the comparison with hydrodynamical simulations (see sec~\ref{sec:simulations}). 

\section{X-ray to radio band SEDs}\label{app:SEDs}
We report here examples of the three cases discussed in section~\ref{sec:importance_radio_sed}, regarding the importance of the radio band in the SED-fitting. The \textit{right} plot is our best fit SED using all the available data, the \textit{left} plot is the same source and photometry but excluding the radio band.

\begin{figure*}
	\includegraphics[width=0.95\textwidth]{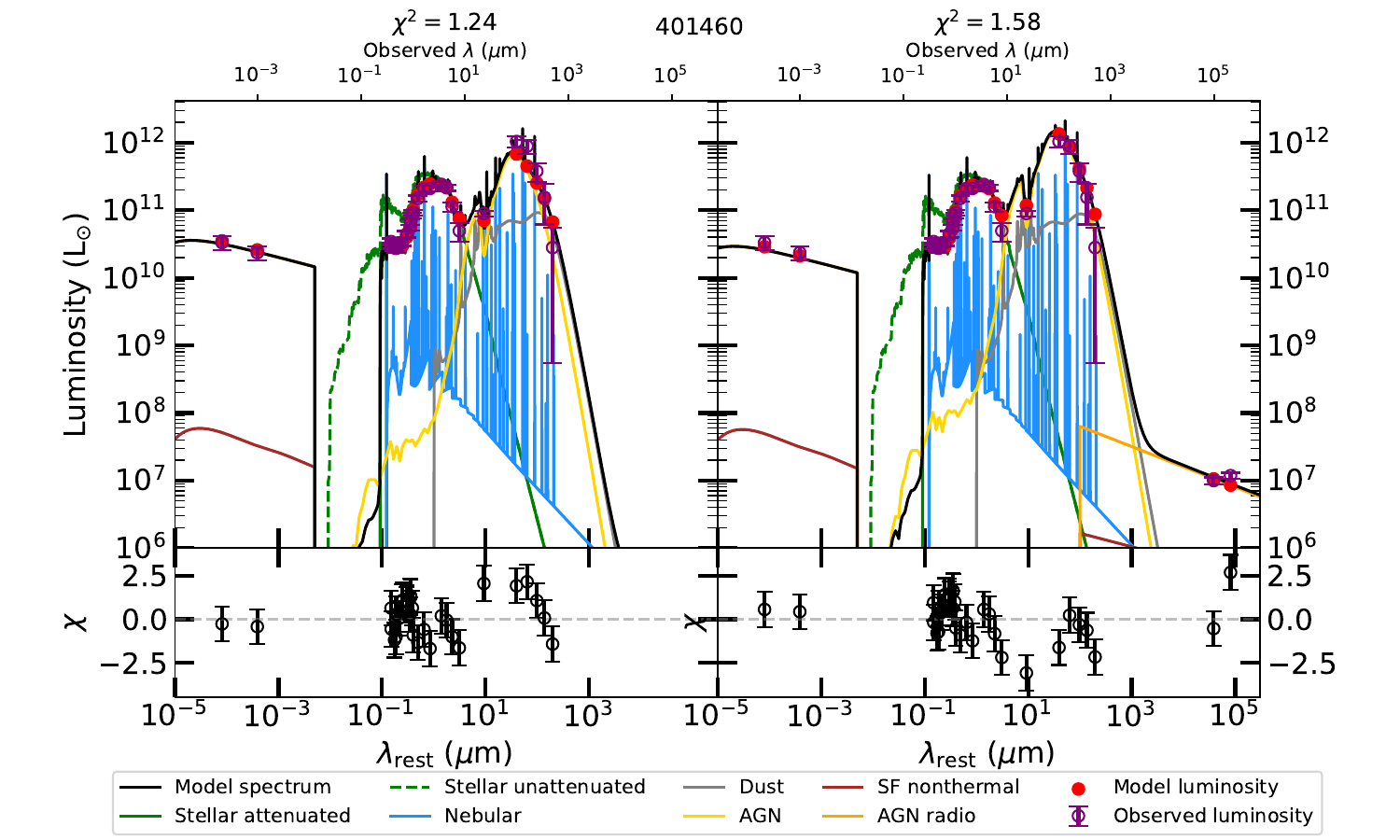}
    \caption{Comparison of the best-fit SED for source zCOSMOS ID401460 with (\textit{left panel}) and without (\textit{right panel}) the inclusion of the radio band photometry. The \textit{purple} points are the observed photometry, the \textit{red} ones refer to the best-fit model luminosity. The \textit{black} line is the best-fit model, composed of the stellar attenuated emission (green continuous line), the nebular emission (\textit{blue}), the dust emission (\textit{grey}), the AGN emission in the optical-IR (\textit{yellow}), the AGN emission at radio and X-ray wavelengths (\textit{orange}), and the SF-related emission at the same bands (\textit{brown}). The \textit{dashed green} line is the intrinsic (i.e. extinction-corrected) stellar emission. For this source, having already a good IR and X-ray coverage, the addition of the radio band to the SED fitting does not significantly change the results.} 
    \label{fig:sed_case1}
\end{figure*}

\begin{figure*}
	\includegraphics[width=0.95\textwidth]{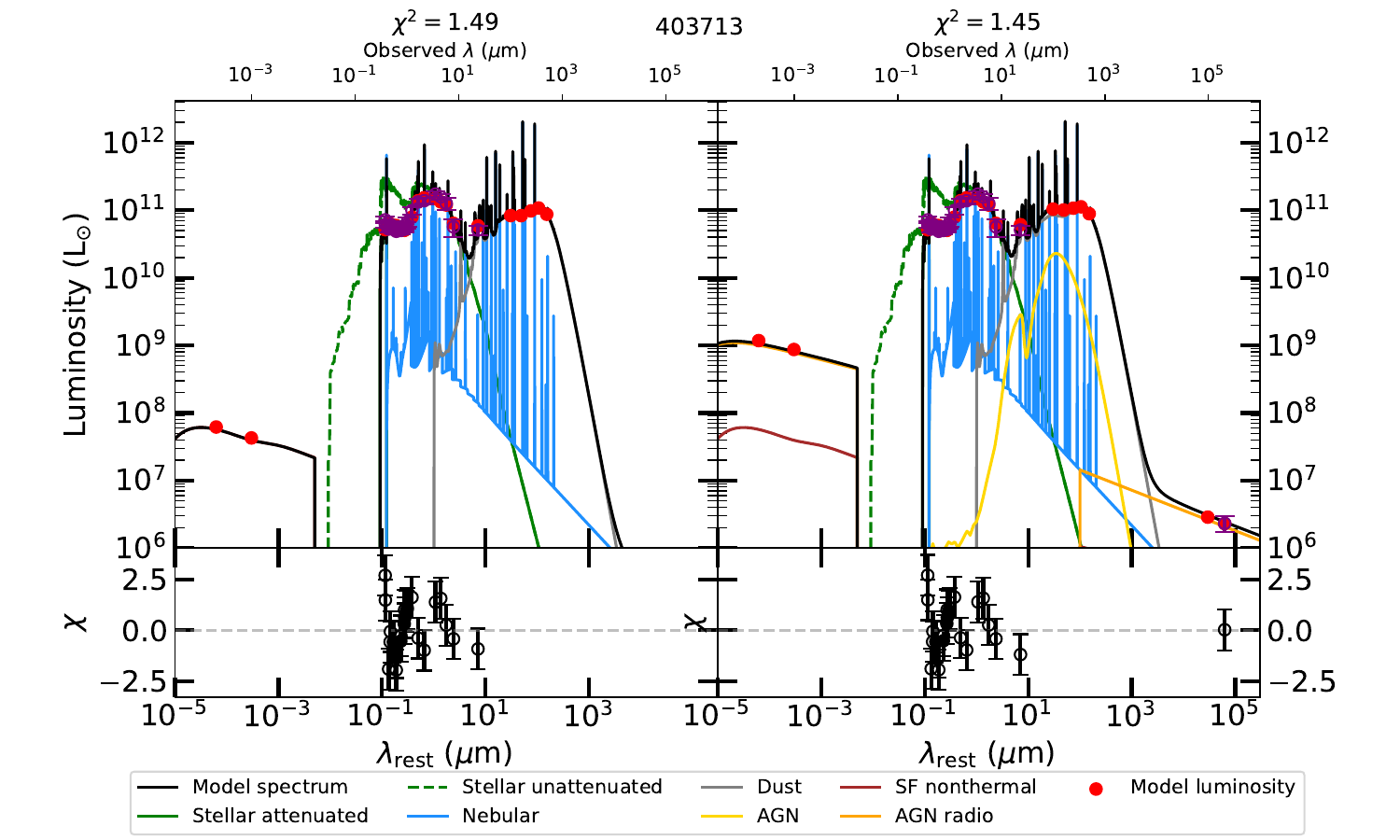}
    \caption{Comparison of the best-fit SED for source zCOSMOS ID403713 with (\textit{left panel}) and without (\textit{right panel}) the inclusion of the radio band photometry. The colour code is the same as in Fig.~\ref{fig:sed_case1}. For this source, we do not have any detection in the X-rays or in the FIR. Without the addition of the radio band, the obscured AGN contribution is not revealed. } 
    \label{fig:sed_case2}
\end{figure*}

\begin{figure*}
	\includegraphics[width=0.95\textwidth]{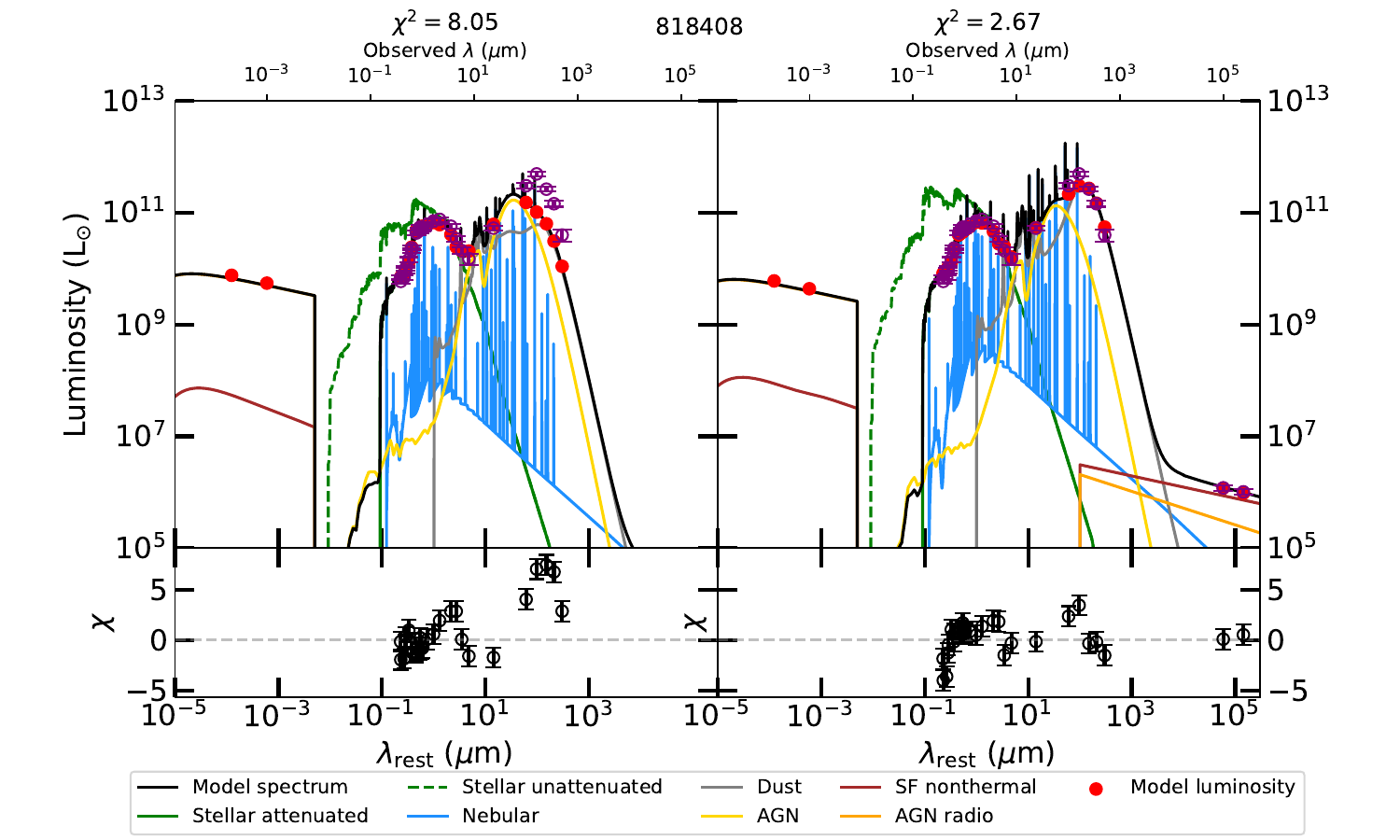}
    \caption{Comparison of the best-fit SED for source zCOSMOS ID818408 with (\textit{left panel}) and without (\textit{right panel}) the inclusion of the radio band photometry. The colour code is the same as in Fig.~\ref{fig:sed_case1}. For this source, we do not have any detection in the X-rays. Without the addition of the radio band, we could not obtain a good fit, especially in the FIR bands, and the contribution of the AGN was overestimated. } 
    \label{fig:sed_case3}
\end{figure*}


\bsp	
\label{lastpage}
\end{document}